\documentclass[aps,twocolumn,amsmath,showkeys,showpacs,prl]{revtex4-1}
\usepackage{dcolumn}
\usepackage{bm}
\usepackage[T1]{fontenc}
\usepackage{amsmath}
\usepackage{graphicx}
\usepackage{tabularx}

\usepackage{graphicx}
\usepackage{dcolumn}
\usepackage{bm}
\usepackage{inputenc}
\usepackage{natbib}
\usepackage{graphicx}  
\usepackage{epsfig}
\usepackage{float}
\usepackage{dcolumn}   
\usepackage{array}
\usepackage{amssymb}   
\usepackage{amsmath}
\usepackage{amstext}
\usepackage{mathtools}
\usepackage{threeparttable}
\usepackage{tablefootnote}
\usepackage{mathrsfs}
\usepackage{xcolor}

\newcommand{\Ket}[1]{\left\vert #1 \right\rangle}

\usepackage [colorlinks=true,allcolors=blue]{hyperref}

\usepackage{hyperref}
\hypersetup{colorlinks=true, linkcolor=blue, citecolor=blue, urlcolor=blue, pdftitle={Rashba in Bi$_2$WO$_6$ Aurivillius phase structure}, pdfauthor={Hania Djani}}

\begin{document}

\title{ Rashba spin-splitting in ferroelectric oxides: from rationalizing to engineering }

\author{Hania Djani$^{1,2,\S}$}
\email{hdjani@cdta.dz}
\author{Andres Camilo Garcia-Castro$^{2,3,\S}$}
\email{acgarciacastro@uliege.be}
\author{Wen-Yi Tong$^{2}$}
\author{Paolo Barone$^{4}$}
\author{Eric Bousquet$^2$}
\author{Silvia Picozzi$^4$}
\author{Philippe Ghosez$^2$}
\email{philippe.ghosez@ulg.ac.be}

\affiliation{$^1$Centre de D\'eveloppement des Technologies Avanc\'ees, cit\'e 20 ao\^ut 1956, Baba Hassen, Alger, Algeria}
\affiliation{$^2$Theoretical Materials Physics, Q-MAT, CESAM, Universit\'e de Li\`ege, B-4000 Li\`ege, Belgium}
\affiliation{$^3$Departamento de F\'isica, Universidad Industrial de Santander, Cra. 27 Cll. 9, Bucaramanga, Colombia}
\affiliation{$^4$Consiglio Nazionale delle Ricerche (CNR-SPIN) c/o Univ. ``G. D'Annunzio'', Chieti, Italy}
\affiliation{\rm ($^\S$These authors contributed equally.)}

\begin{abstract}

Ferroelectric Rashba semiconductors (FERSC), in which Rashba spin-splitting can be controlled and reversed by an electric field, have recently emerged as a new class of functional materials useful for spintronic applications. The development of concrete devices based on such materials is, however, still hampered by the lack of robust FERSC compounds. Here, we show that the coexistence of large spontaneous polarisation and sizeable spin-orbit coupling is not sufficient to have strong Rashba effects and clarify why simple ferroelectric oxide perovskites with transition metal at the B-site are typically not suitable FERSC candidates. By rationalizing how this limitation can be by-passed through band engineering of the electronic structure in layered perovskites, we identify the Bi$_2$WO$_6$ Aurivillius crystal as the first robust ferroelectric with large and reversible Rashba spin-splitting, that can even be substantially doped without losing its ferroelectric properties. Importantly, we highlight that a unidirectional spin-orbit field arises in layered Bi$_2$WO$_6$, resulting in a protection against spin-decoherence.
We highlight moreover that a unidirectional spin-orbit field arises in Bi$_2$WO$_6$, in which the spin-texture is so protected against spin-decoherence.
\end{abstract}

\maketitle

 \section{I. Introduction}

In non-magnetic solids, one can naively expect the energy bands of electrons of up and down spins to be degenerate in absence of magnetic fields. However, in systems that break spatial inversion symmetry, \textit{e.g.} at surfaces and interfaces but also in non-centrosymmetric bulk crystals
, spin-orbit coupling (SOC) can lift such spin band degeneracy through the so-called Rashba and Dresselhaus effects \cite{PhysRev.100.580, rashba1960, rashba1984}. During the last decade, 
these phenomena have attracted  increasing interests in various fields, including spintronics, quantum computing, topological matter and cold atom systems \cite{Manchon2015,Varignon18}. 

Recently, the concept of ferroelectric Rashba semi-conductors (FERSC) has been introduced \cite{10.3389/fphy.2014.00010}. It defines a new class of functional materials combining ferroelectric and Rashba effects, in which the spin texture related to the Rashba Spin Splitting (RSS) can be electrically switched upon reversal of the ferroelectric polarisation. As such, FERSC offer exciting perspectives for spintronic applications. The Rashba spin precession of a current injected in such materials can be controlled in a non-volatile way by their reversible ferroelectric  polarisation.  
Moreover, FERSC allow to envision new devices interconverting electron- and spin-currents based-on the Edelstein \cite{EDELSTEIN1990233} and reverse-Edelstein \cite{Sanchez2013} effects.  In two-dimensional  ferroelectric materials with in-plane polarization and strong anisotropy in the electronic structure, the spin-orbit field (SOF) was proposed to have unidirectional out-of-plane alignment:
$\vec{\Omega}_{SOF}(\vec{k})=\alpha(\vec{P} \times \vec{k}) = \alpha k_y \hat{z}$, where $\alpha$ is a system-dependent coefficient \cite{PSH.FE.preprint}. In such a case, injected electrons with in-plane spins would therefore precess around the $\vec{z}$ axis, giving rise to a long-lived persistent spin helix (PSH), a concept originally proposed for quantum-wells of III-V semiconductors with fine-tuned Dresselhaus and Rashba coefficients \cite{PSH.Loss2003,PSH.Bernevig2006,PSH.Schliemann2016,PSH.nature2009,PSH.nphys2009,PSH.RMP2017} and very recently extended to a subclass of noncentrosymmetric bulk materials \cite{PSH.Tsymbal2018}. Independently, FERSC can also, in some cases, exhibit ferro-valley properties \cite{Tong2016}.

The basic idea of FERSC was first put forward theoretically in bulk GeTe \cite{picozzi1} and then  experimentally confirmed in GeTe thin films \cite{picozzi1,ADMA:ADMA201503459,Rinaldi2018}. Unfortunately, GeTe does not appear as the best candidate for concrete applications, due to its very small bandgap and related large leakage currents that, in most cases, prevent polarization switching \cite{10.3389/fphy.2014.00010}. The identification of alternative robust FERSC is therefore mandatory to achieve full exploitation of the concept. Although different directions have been explored \cite{davanse,Stroppa14,PhysRevLett.115.037602,Narayan15,Zhong15,DiSante16,Zhang17,VarignonU}, no really convincing candidate has emerged yet. 

Here, we rationalise by means of first-principles approaches (see Methods) the discovery of a promising FERSC in the family of oxide perovskite compounds. Focusing first on simple perovskites, we highlight that robust ferroelectricity and SOC are necessary but not sufficient conditions to get an efficient FERSC. Furthermore, we clarify why these materials are typically not suitable candidates. We then propose a strategy to by-pass their intrinsic limitation in layered perovskites and identify the Bi$_2$WO$_6$ Aurivillius phase as the first robust ferroelectric with large and reversible Rashba spin-splitting at the bottom of the conduction band and unidirectional SOF. We finally show that a significant $n$-type doping does not lead to a loss of its ferroelectric properties, suggesting the possibility of creating a doped FERSC appropriate for practical applications. 
\section{II. Method}
First-principles calculations relied on Density Functional Theory (DFT) \cite{PhysRev.136.B864,PhysRev.140.A1133} within the Projected Augmented Waves (PAW) method \cite{Blochl1994} as implemented in the Vienna $Ab$ $initio$ Simulation Package (\textsc{vasp}) \cite{Kresse1996,Kresse1999}. Many results were also checked using ABINIT \cite{Gonze2002,Gonze2005,Gonze2009} with norm-conserving pseudopotentials.  
The exchange-correlation effects were estimated within the Generalized Gradient Approximation (GGA) using the PBEsol parameterization \cite{Perdew2008}. The following electrons were treated as valence states: Bi(5$d^{10}$6$s^2$6$p^3$), W(5$p^6$6$s^2$5$d^4$), and O(2$s^2$2$p^4$). 
Convergence was reached using a Monkhorst-Pack \cite{PhysRevB.13.5188} 8$\times$8$\times$2 \emph{k}-point mesh for Bi$_2$WO$_6$ (8$\times$8$\times$8  for WO$_3$) and a 600 eV energy cutoff. Structural relaxations were converged until forces are less than 1 meV$\cdot$\r{A}$^{-1}$. The spin-orbit coupling was included into the calculations as in Ref. \onlinecite{Hobbs2000}. 
Electron doping was performed by adding electrons to the total electronic density and introducing a neutralizing homogeneous background charge to compensate the additional electrons, as previously done in several works \cite{PhysRevB.97.054107, PhysRevLett.109.247601, C5TC03856A, PhysRevB.86.214103}; electronic and atomic relaxations were carried out at fixed volume \cite{Bruneval15}. The spin-texture was analyzed using the script PyProcar \cite{pyprocar}, the structural distortion was analysed with \textsc{amplimode} \cite{amplimode} and the figures of atomic structures were elaborated with \textsc{vesta} \cite{vesta}.\\

\section {III. Results}

\subsection{A. Simple perovskites}
Ideal FERSC materials must meet a series of requirements. They should be non-magnetic ferroelectrics insulators with a sizable switchable polarisation and a reasonable bandgap. They should include heavy ions with large SOC exhibiting a significant RSS close to the valence or conduction band edge, which should be reversible with the polarization and, for applications based on spin/charge currents, should survive to appropriate doping. 

Regarding ferroelectricity, it is natural to look to $d^0$ $AB$O$_3$ perovskites with a transition metal at the B-site \cite{Hill}, in which the bandgap is formally between O-$2p$ and $B$-$d$ states. As such, a large RSS around the bandgap would be more easily achieved by means of a heavy cation at the $B$-site while B-type ferroelectricity would likely favor an efficient polarization control of the RSS. 

\begin{figure}[!t]
 \centering
 \includegraphics[width=9cm,keepaspectratio=true]{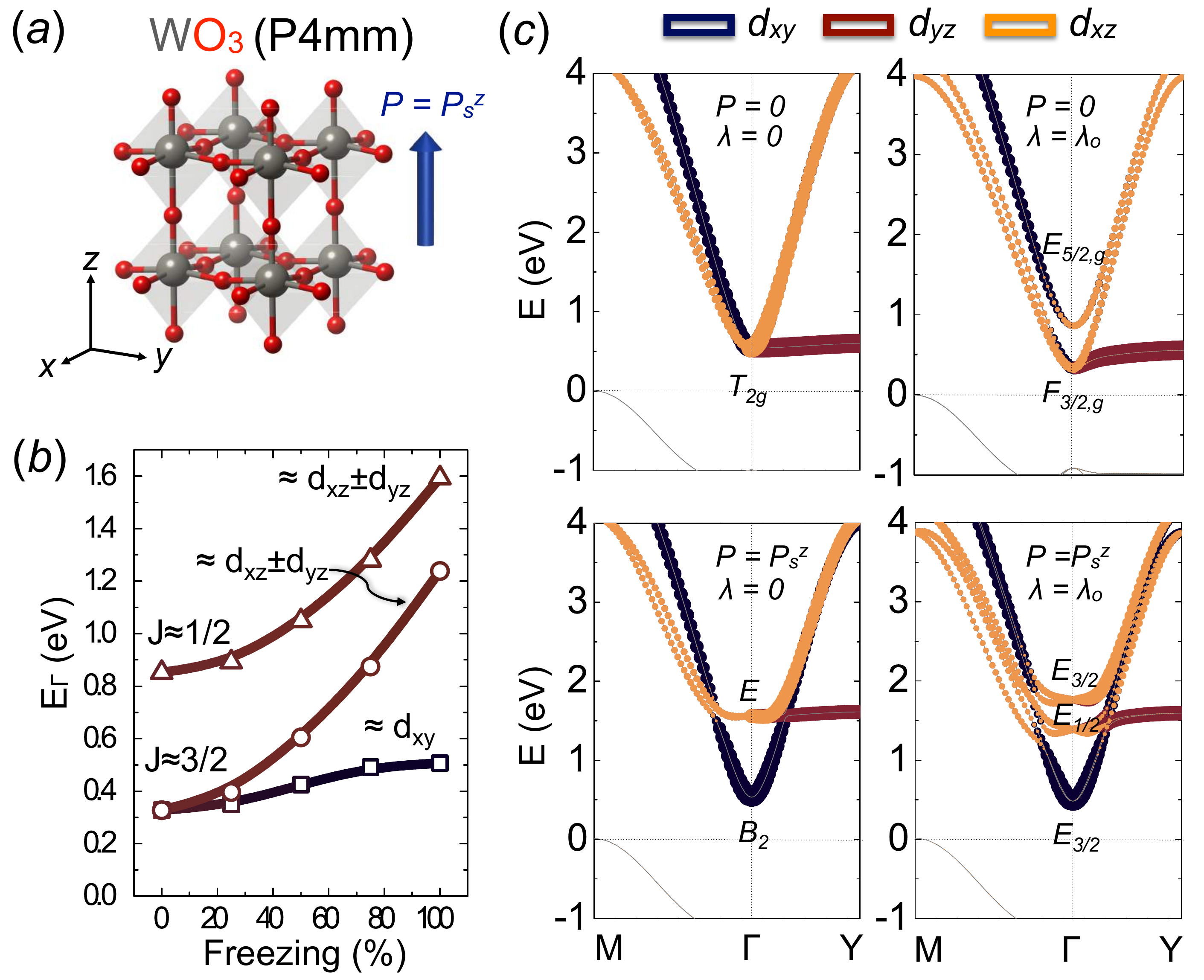}
  \caption{Electronic dispersion curves in the $P4mm$ phase of WO$_3$. ($a$) Sketch of the $P4mm$ phase of WO$_3$, with $P_s$ along the $z$-axis. ($b$) Evolution of the electronic band structure around the Fermi level when activating the polar distortion ($P=0 \rightarrow P_s^z$) and SOC ($\lambda = 0 \rightarrow \lambda_0$). Projection on the $t_{2g}$ orbitals ($d_{xy}$, $d_{yz}$, $d_{zx}$) of the reference structure ($P=0$, $\lambda = 0$) are highlighted in colors. ($c$) Evolution of the splitting of the original $t_{2g}$ states at $\Gamma$-point for increasing polar distortion ($P=0 \rightarrow P_s^z$) when including SOC ($\lambda = \lambda_0$). The projection on the $t_{2g}$ orbitals are highlighted by mixing colors as in panel ($c$).   } 
\label{fig:1}
\end{figure}

Tungsten oxide, WO$_3$, is in line with previous requirements. It adopts the perovskite structure with an empty $A$-site and a heavy W atom on the $B$-site (see Fig. \ref{fig:1}(a)).  It is also an insulator with formal $d^0$ occupancy of the W 5$d$ states.  Although not intrinsically ferroelectric -- it adopts a nonpolar $P2_1/c$ ground state \cite{salje} --, a recent study highlighted that it possesses low-energy metastable ferroelectric phases with large spontaneous polarizations ($P_s \approx 50-70 \; \mu$C$\cdot$cm$^{-2}$) arising from the opposite motion of W and O atoms (Supplemental Material I.A) \cite{hamdi}. Although never observed experimentally, these polar phases appear to be relevant prototypical states to investigate and rationalize the interplay between polarization and SOC in perovskite-like systems.

Fig. \ref{fig:1}(a) presents a sketch of the $P4mm$ ferroelectric phase of WO$_3$, which exhibits a spontaneous polarization along the cartesian $z$-axis ($P_s^z = 54$ $\mu$C$\cdot$cm$^{-2}$). 
In Fig. \ref{fig:1}(b), we show the calculated electronic band structure around the band gap of the cubic and tetragonal $P4mm$ phase of WO$_3$ with and without SOC.
In the cubic phase ($P_s^z=0$) without SOC ($\lambda = 0$), the bottom of the conduction band of WO$_3$ is at $\Gamma$ and consists of triply-degenerate state of $t_{2g}$ symmetry (pure $d_{xy}$,$d_{yz}$ and $d_{zx}$ orbitals). On the one hand, activating SOC ($\lambda = \lambda_0$) mixes the three $t_{2g}$ states and produces a splitting $\Delta_{SOC}$ between a doubly-degenerate low-energy state of $F_{3/2,g}$ symmetry ($J=3/2$) and a higher-energy state of $E_{5/2,g}$ symmetry ($J=1/2$) \cite{PhysRev.171.466}. 
On the other hand, the $P4mm$ phase ($P=P_s^z$) without SOC has a splitting $\Delta_{FE}$ between a low-energy state of $B_2$ symmetry (pure $d_{xy}$ orbital perpendicular to $P_s^z$ at first perturbative order) and a higher-energy doubly-degenerate state of $E$ symmetry (mixed $d_{yz}$ and $d_{zx}$ orbitals, partly hybridized with O $2p$) \cite{Wolfram2006}. 
In the presence of both SOC and ferroelectric polarization, three distinct levels of $E_{3/2}$, $E_{1/2}$ and $E_{3/2}$ symmetry are present. 
For small amplitude of $P_s^z$, $\Delta_{FE}$ is small compared to $\Delta_{SOC}$ and all the three levels arise from a mixing of the three $t_{2g}$ orbitals (see Fig.\ref{fig:1}(c)). 
As $P_s^z$ and $\Delta_{FE}$ increase, the lowest $E_{3/2}$ acquires a dominant $d_{xy}$ character (like the $B_2$ state without SOC) while the higher-energy $E_{1/2}$ and $E_{3/2}$ levels are a mixing of $d_{yz}$ and $d_{zx}$ orbitals. 
This is supported by a simple tight-binding model (see Supplemental Material  I.B). 

Estimate of the RSS strength in the $P4mm$ phase through the effective Rashba parameter $\alpha_R = 2 E_R/k_R$ \cite{Winkler2003,ishizaka} gives a sizable value $\alpha_R \approx 0.7$ eV$\cdot$\r{A} for the upper bands linked to $E_{3/2}$ and $E_{1/2}$ states.  However, $\alpha_R \approx 0$ for the band linked to the lowest $E_{1/2}$ state with strongly dominant $d_{xy}$ character ($d_{xy}$ is perpendicular to $P_s^z$). 

\begin{figure*}[t!]
 \centering
 \includegraphics[width=15.0cm,keepaspectratio=true]{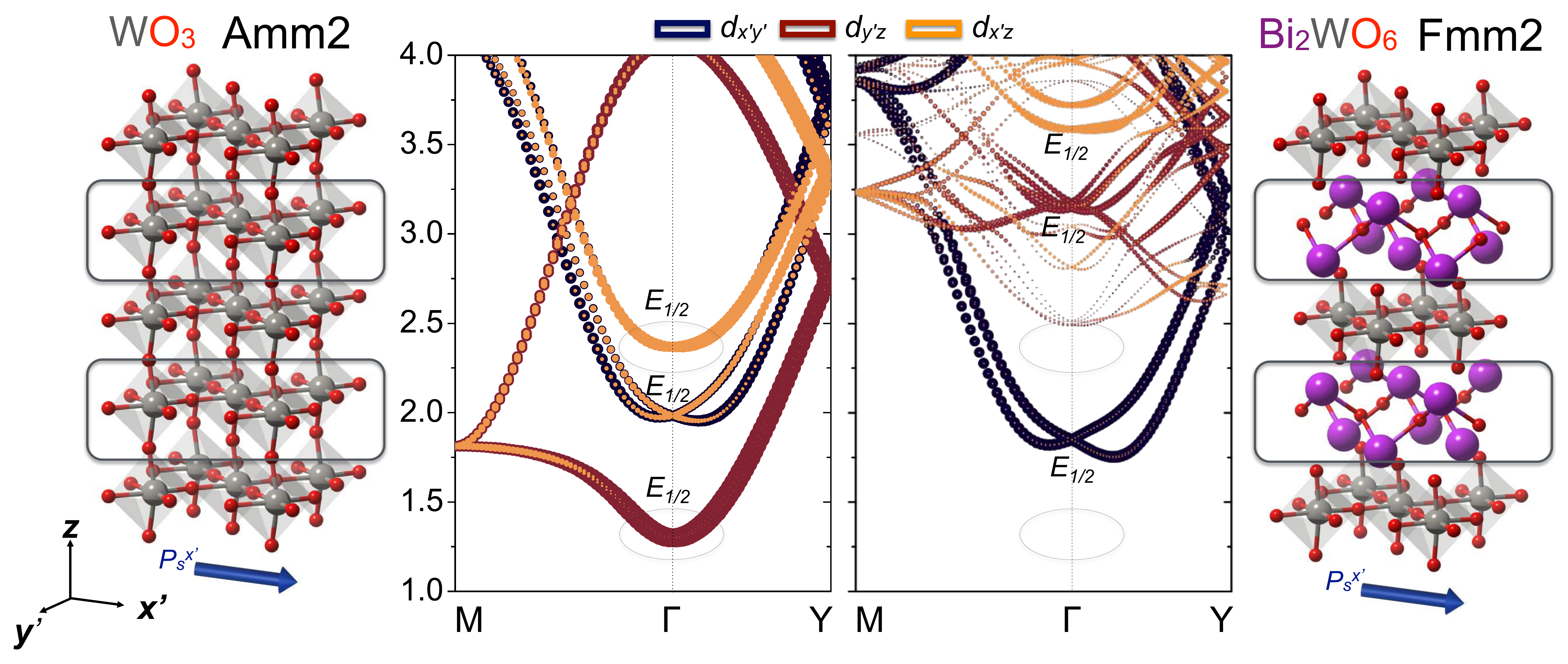}
\caption{Comparison of WO$_3$ and Bi$_2$WO$_6$. Sketch of the atomic structure and electronic dispersion along the $\Gamma$--Y direction focusing on the lowest conduction states, $E_F=0$) of the $Amm2$ phase of WO$_3$ (left) and $Fmm2$ phase of Bi$_2$WO$_6$ (right). Contribution of the $t_{2g}$ orbitals ($d_{xy}$, $d_{yz}$, $d_{zx}$) to the different electronic states are highlighted in colors.}
  \label{fig:2}
\end{figure*}

The same conclusions apply to the ferroelectric $Amm2$ phase of WO$_3$ (see Fig. \ref{fig:2} and Supplemental I.C) where the polarization is along the $xy$ pseudo-cubic direction ($x'$ in a reference axis rotated by 45$^o$ around $z$ with respect to $x$) and with a calculated $P_s^{x'} = 69$ $\mu$C cm$^{-2}$ . In this orthorhombic phase, the reference $t_{2g}$ states are split in three levels of $E_{1/2}$ symmetry.  The lowest state has a strongly dominant $d_{y'z}$ character ($d_{y'z}$ is perpendicular to $P_s^{x'}$) and does not show any significant RSS. 

These results are generic to ABO$_3$ perovskites and remain valid in presence of a (``non-empty'') A-cation, as in KTaO$_3$ (see Ref. \cite{Tao2016} and Supplemental II): the first unoccupied $d$-band does not show RSS in the presence of ferroelectric polarization. 

A natural question at this stage is why the lowest $t_{2g}$ state does not show RSS. As highlighted from a simple tight-binding model restricted to the $t_{2g}$ subspace (see Supporting Information  I. B), all the three levels are allowed to show RSS but $\alpha \propto \Delta_{SOC}/\Delta_{FE}$ and should vanish for all states in the limit of large $\Delta_{FE}$. The question is then rather  why the upper $t_{2g}$ states show significant RSS. A plausible explanation is their interactions with the 2p states of bridging oxygen atoms. Combining an extended tight-binding model and first-principles calculations, we instead demonstrate  that the dominant effect comes actually from their hybridization with the $e_g$ states. 

This rationalize that significant RSS can appear in the $t_{2g}$ conduction states of $d^0$ $AB$O$_3$ perovskites with heavy $B$-site atoms. 
However, RSS is restricted to the upper $t_{2g}$ levels showing significant hybridization with the $e_g$ states. Consequently, achieving a large $\alpha_R$ at the conduction band bottom of perovskites would require to get rid of the lowest energy state associated with the $d_{\perp}$ orbital perpendicular to $P_s$. 
As we now show, this can be achieved if one confines the ferroelectric material in the direction perpendicular to $P_s$, which is naturally realized for WO$_3$ in the Bi$_2$W$_n$O$_{3n+3}$ Aurivillius series, a family of single-phase layered compounds alternating WO$_3$ perovskite blocks with Bi$_2$O$_2$ fluorite-like layers.


\subsection{B. Layered perovskites }

Bi$_2$WO$_6$ is the $n=1$ member of the Bi$_2$W$_n$O$_{3n+3}$ series. 
It is a strong ferroelectric with large polarization ($P_s \approx 50$ $\mu$C cm$^{-2}$) and high Curie temperature ($T_c = 950 $ K).
It has a measured experimental gap of 2.7 - 2.8 eV \cite{Zhou2015, Lv2016} defined between the O $2p$ and W $5d$ states of the perovskite block (see Supplemental Material III.B). Furthermore, Bi$_2$WO$_6$ is prone to n-type doping.\cite{Noguchi06,Noguchi07}

Bi$_2$WO$_6$ exhibits a polar orthorhombic $P2_1ab$ phase up to 670$^o$C, at which it undergoes a phase transition to another polar orthorhombic phase of  $B2cb$ symmetry, stable up to 950$^o$C \cite{mcDowell,djani}. As discussed in Ref.\cite{djani}, the polar $B2cb$ and $P2_1ab$ phases are small distortions of the same reference $I4/mmm$ high-symmetry structure and arise from the consecutive condensation of independent atomic motions: (i) a polar distortion  along the $x'$-axis ($\Gamma_5^{-}$ symmetry) lowering the symmetry from $I4/mmm$ to $Fmm2$, (ii) tilts of the oxygen octahedra along the $x'$-axis ($X_3^{+}$ symmetry) lowering further the symmetry to $B2cb$ and (iii) rotations of the oxygen octahedra around the $z$-axis ($X_2^{+}$ symmetry) bringing the system in its $P2_1ab$ ground state.

The polar $Fmm2$ phase of Bi$_2$WO$_6$ is comparable to the $Amm2$ phase of bulk WO$_3$ (Fig. \ref{fig:2}) with a spontaneous polarization $P_s^{x'}$ in the $xy$ pseudo-cubic directions and oriented in plane (\textit{i.e.} perpendicular to the stacking direction). 
In Fig. \ref{fig:2} we compare the electronic band structure of $Amm2$ WO$_3$ and  $Fmm2$ Bi$_2$WO$_6$ in the presence of SOC. 
In both cases, the $t_{2g}$ states at $\Gamma$ are split into 3 distinct $E_{1/2}$ levels. 
However, in Bi$_2$WO$_6$ due to the asymmetry imposed by the Bi$_2$O$_2$ layers along the $z$-axis, the states associated to the W $d_{x'z}$ and $d_{y'z}$ orbitals are pushed to much higher energy than the $d_{x'y'}$. 
Consequently, the $E_{1/2}$ level at the conduction band bottom is now the one with  dominant $d_{x'y'}$ character and it exhibits a large $\alpha_R$ of 1.28 eV$\cdot$\r{A}.

\begin{table}[!b]
\caption{Spontaneous polarization ($P_s$), $k$-vector splitting ($k_R$), energy splitting ($E_R$), Rashba parameter ($\alpha_R$), and theoretical energy gap (E$_{g}^{DFT}$) for distinct ferroelectric phases of Bi$_2$WO$_6$ and few selected reference systems.}\label{tab:1}
\centering
\small
\begin{tabular}{lcccccc}
\hline
\hline
&  &P$_s$  &$k_R$  & $E_R$ &$\alpha_R$   & E$_{g}^{DFT}$    \\
&  &($\mu$C$\cdot$cm$^{-2}$)  &(\r{A}$^{-1}$) &(meV) &(eV$\cdot$\r{A})  & (eV)  \\
\hline
Bi$_2$WO$_6$ & $Fmm2$ &78   &  0.155     &  99.4     & 1.28    & 1.82    \\
                          & $B2cb$   &67   &  0.136     &  53.0     &  0.78  & 1.77  \\
                          & $B2cm$   &68  &  0.168     & 101.9   &  1.22  &  1.94 \\
                          &$P2_1ab$ &65   & 0.163 & 71.4   &0.88 & 1.83 \\
BiAlO$_3$ \cite{davanse}  &$R3c$    & 79  & 0.04   & 7        &0.39   & 2.57  \\
GeTe \cite{picozzi1} &$R3m$      & 60  & 0.09   & 227    &4.80   & 0.38 \\
BiTeI \cite{ishizaka}  & $P3m1$      & - &  0.052 & 100    &3.85   & 0.43 \\
\hline
\hline
\end{tabular}
\end{table}

Since the $Fmm2$ phase is not observed experimentally, we now analyze how oxygen octahedra rotations ($X_3^{+}$ and $X_2^{+}$) present in the $B2cb$ and $P2_1ab$ phase on top of the polar distortions ($\Gamma_5^{-}$) affect the RSS.
In order to clarify the independent role of $X_3^{+}$ and $X_2^{+}$ distortions,  we compare, in Table \ref{tab:1}, $\alpha_R$ in distinct fully relaxed ferroelectric phases: $Fmm2$ ($\Gamma_5^{-}$), $B2cb$ ($\Gamma_5^{-}$+$X_3^{+}$), $B2cm$ ($\Gamma_5^{-}$+$X_2^{+}$) and $P2_1ab$ ($\Gamma_5^{-}$+$X_3^{+}$+$X_2^{+}$). 
It appears that the RSS is dominantly produced by the polar $\Gamma_5^-$ distortion, while oxygen rotations play a detrimental but much minor role (see Supporting Information III.C): the $X_3^{+}$ distortion tends to decrease $\alpha_R$, while the $X_2^{+}$ distortion has no direct effect. 
In fact $k_R$ stays almost unchanged in all the phases, while $E_R$ is more affected.  Overall,  the amplitude of $\alpha_R$ in the $P2_1ab$ ground state is slightly reduced but remains comparable to that of the $Fmm2$ phase. 

\begin{figure*}[t!]
 \centering
 \includegraphics[width=15.0cm,keepaspectratio=true]{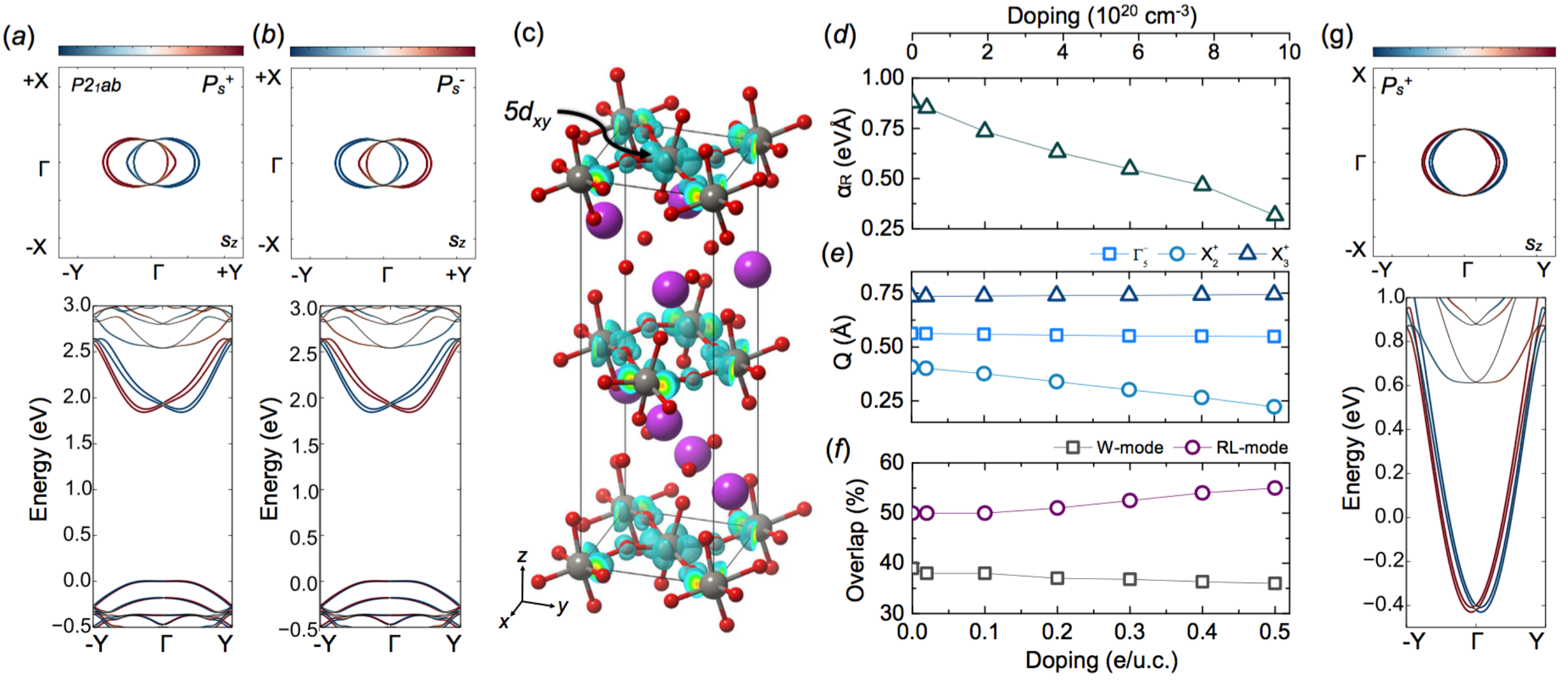}
 \caption{Pristine and n-doped P2$_1$ab  phase of Bi$_2$WO$_6$. (a-b) $S_z$ spin-projected band dispersion around the Fermi level (bottom) and spin-texture into $k_xk_y$ plane at E = 2.0 eV (top) of pristine  observed at Bi$_2$WO$_6$  in its P2$_1$ab  phase for (a) up and (b) down polarization directions. A reversible spin splitting is observed at the conduction band bottom. (c) Charge-density of the conduction bands in the P2$_1$ab  phase of Bi$_2$WO$_6$ with a doping of 0.5 $e$/u.c.. (d-f) Evolution with the electronic doping concentration of respectively (d) the  Rashba parameter, (e) the $\Gamma_5^-$, $X_2'$ and $X_3'$ distortion amplitudes and (f) the respective W-mode and RL-mode contributions to $\Gamma_5^-$ distortion in the P2$_1$ab phase of Bi$_2$WO$_6$. (g) $S_z$ spin-projected bands (bottom) and spin-texture into $k_xk_y$ plane at E$_F$=0 eV (top) in the P2$_1$ab  phase of Bi$_2$WO$_6$ with a doping of 0.5 $e$/u.c.. A characteristic 2DEG-like behavior is observed from the parabolic band shape.}
 \label{fig:3}
\end{figure*}

Fig. \ref{fig:3}(a) shows the electronic dispersion curves of the $P2_1ab$ phase, highlighting the significant spin splitting at the conduction band bottom. We notice an additional band splitting due to the presence of the oxygen tilts ($X_3^{+}$ distortion) that double the unit cell in the $y'z$-plane.
Constant energy maps are also shown for an energy of 2.0 eV, along with the corresponding spin texture. The relative orientation of the coupled $k$ and $S$ components is determined by the symmetry of the system; in our case, the four polar phases belong to the $C_{2v}$ point group that contains a $C_{2x}$ two-fold rotation around the polar $x'$-axis and two mirror planes, $m_{\bot y}$ and $m_{\bot z}$. The electronic structure has the shape of two partially overlapping revolution paraboloids with revolution axes symmetrically shifted in opposite directions with respect to $k_y$ = 0. These two paraboloids are associated to electrons with opposite $S_z$ spin component and an additional $S_y$ contribution ensuring rotation of S in the region where the ellipsoids cross. No $S_x$ component is observed, consistently with the Rashba-like effect. The RSS is proportional to the polarization and is reversed under polarization switching (Fig. \ref{fig:3}(b)). 

It can also be noted that the spin splitting vanishes along the $\Gamma$ $\rightarrow$ \textit{X} path, corresponding to the polarization direction. As such, all the symmetry constraints and design criteria proposed in Ref. \cite{PSH.FE.preprint,PSH.Tsymbal2018} in order to have a unidirectional SOF are met. We therefore conjecture  the spin-lifetime in Bi$_2$WO$_6$ to be long, due to reduction of spin decoherence mechanisms  (the latter being related in Bi$_2$WO$_6$ only to higher order momentum k-cube term in  SOF).  In addition, we expect a long-lived and nanometer-sized PSH, which could be of high relevance for future spintronic applications. 

\subsection{C. Doping}
So far, we have shown Bi$_2$WO$_6$ to be a robust switchable ferroelectric with large reversible RSS at the conduction band bottom. 
To be also of practical utility for spintronic applications based on charge/spin currents, it should additionally be possible to dope it with electrons, which contrary to some other Aurivillius, appears to be naturally the case \cite{Noguchi06,Noguchi07}.  Moreover, it should keep its FERSC properties when n-doped. This is far from obvious, since adding conduction electrons is expected to suppress ferroelectricity (and related RSS). Nevertheless, recent studies have shown that prototypical ferroelctrics like BaTiO$_3$ can preserve their ferroelectric distortion under n-doping concentrations up to 0.1$e$/u.c. \cite{PhysRevLett.104.147602, PhysRevLett.109.247601}.

In Fig. \ref{fig:3}, we report the evolution of structural and electronic properties of the $P2_1ab$ phase of Bi$_2$WO$_6$ under electron doping (see Methods). In line with the electronic structure of the pristine material, doping electrons occupy the W $5d$ states around the conduction band bottom. Due to the dominant $d_{x'y'}$ character of these states, these electrons form a two-dimensional electron gas (2DEG) confined in the perovskite layer (Fig. \ref{fig:3}(c)). Amazingly, symmetry-adapted mode analysis of the atomic distortion of the doped structure with respect to the $I4/mmm$ reference structure indicates that the global $\Gamma_5^-$ polar distortion remains constant under electron doping (Figure \ref{fig:3}e), rather than being suppressed. Further insights are given by the projection of this distortion on the phonon eigendisplacement vectors of the $I4/mmm$ reference (in Fig. \ref{fig:3}(f)). The global $\Gamma_5^-$ polar distortion arises in fact  from the condensation of two distinct phonon modes: a ``W-mode'' confined in the WO$_3$ layer and related to the off-centering of W in its O octahedron cage and a ``RL-mode'' ($i.e.$ rigid-layer mode \cite{machado, djani}), related to a nearly rigid motion of the Bi$_2$O$_2$ layer with respect to the perovskite block. 
Although the global polar distortion remains constant under n-doping, the contribution of the W-mode is progressively suppressed when increasing the population of the W 5$d_{xy}$ states, while that of the RL-mode is amplified.

Concomitantly with the suppression of the W-mode distortion, $\alpha_R$ (Fig. \ref{fig:3}(b)) is progressively reduced under doping, highlighting that  large polar distortion is not enough to lead to large $\alpha_R$; rather, the polar distortion pattern must occur around the W atom responsible for the RSS, as in the W-mode. 
Although progressively reduced, $\alpha_R$ keeps nevertheless a sizable value up to large n-doping:  at a doping level of 0.5 $e^-$/u.c. ($\approx 10^{21}$ cm$^{-3}$), $\alpha_R$ is still as large as 0.3 eV$\cdot$\r{A}. Fig.\ref{fig:3}(g) shows the related electronic dispersion curves and spin texture. 

\section{IV. Conclusion}

Combining first-principles calculations, symmetry analysis and tight-binding models, we have first rationalized step by step the RSS in the important family of ABO$_3$ perovskites with a transition metal at the $B$-site, demonstrating why they typically do not show significant RSS at the conduction band bottom. Relying on the concept of band-structure engineering in layered structures, we have then identified the Aurivillius Bi$_2$WO$_6$ compound to be the first known ferroelectric oxide to show a large Rasba-like spin splitting at the conduction band bottom that can be reversed upon application of an external electrical field. Beyond being a practical ferroelectric, Bi$_2$WO$_6$ offers additional and appealing peculiarities with respect to previously proposed FERSC candidates: {\em i}) a unidirectional spin-orbit field (arising from the combined presence of in-plane polarization, strong layering-induced anisotropy in the electronic structure and related symmetry properties) that protects the spin-texture from spin dephasing; {\em ii}) the persistence of desired properties (such as robust ferroelectricity, large Rashba spin splitting and unidirectional spin-orbit field) upon sizable n-doping.  

A similar behavior can {\it a priori} be found in other ferroelectric Aurivillius phases, like SrBi$_2$Ta$_2$O$_9$, or even Bi$_4$Ti$_3$O$_{12}$. 
However, the RSS depends on the strength of the $B$-cation SOC that increases with the oxidation state \cite{JCC:JCC21011} and $P_s$, which are maximized in W-based compounds (see Supplemental Material IV). 
Within the Bi$_2$W$_n$O$_{3n+3}$ series, Bi$_2$W$_2$O$_9$ and Bi$_2$W$_3$O$_{12}$ shows a large $\alpha_R$ as well (Supplemental IV). However, Bi$_2$W$_2$O$_9$ is not ferroelectric \cite{Champarnaud99} and Bi$_2$W$_3$O$_{12}$ has not been synthesized yet. Therefore,  Bi$_2$WO$_6$ emerges as the best candidate so far for large RSS and unidirectional SOF in the whole family of perovskite-based oxides, calling for experimental confirmations of our theoretical predictions. Our work also motivate and rationalize the search of alternative candidates in other families of naturally layered perovskites like Ruddlesden Popper and Dion-Jacobson series \cite{Benedek15}.

\textbf{Acknowledgements}\\
Work supported by F.R.S.-FNRS project HiT4FiT, ARC project AIMED and M-ERA.NET project SIOX. Computational resources provided by the Consortium des Equipements de Calcul Intensif (CECI), funded by the F.R.S.-FNRS under the Grant No. 2.5020.11 and the Tier-1 supercomputer of the F\'ed\'eration Wallonie-Bruxelles funded by the Walloon Region under the Grant No 1117545. EB thanks the FRS-FNRS. H.D. and Ph.G. acknowledge support from Algerian-WBI bilateral cooperative project.   \\

\textbf{Author contributions}\\
PhG conceived the study with H.D., A.C.G.C. and E.B. and supervised the work. H.D. and A.C.G.C. did the first-principles calculations and analysed the results. P.B., S.P., W.Y.T. and Ph.G. interpreted the electronic band structures and rationalized the RSS. Ph.G. wrote the manuscript with H.D., A.C.G.C. from inputs of all authors. All authors discussed the results and commented on the manuscript. H.D. and A.C.G.C.contributed equally to this work.

\onecolumngrid
\appendix

\section{Supplemental Material}

\section{I. The case of WO$_3$}

\subsection{A. Relaxed polar structures}

In TABLE S \ref{wo3}, we report the main features of the low energy metastable $P4mm$ and $Amm2$ polar phases of WO$_3$ (more information about the other phases of WO$_3$ and their internal energy can be found in  Ref. \onlinecite{hamdi}). We notice that the theoretical band gap E$_g$ is much larger in the $Amm2$ phase than in the $P4mm$. This behavior, observed in several perovskites, is explained in term of $B$-cation off-centering displacements that increases the anti-bonding character of the orbital at the CBM  \cite{rappe}.

\begin{table*}[b!]
\caption{Optimized lattice parameters, electronic gap, E$_g$, and spontaneous polarization, P$_s$, of the different phases of WO$_3$ as computed including with SOC.} 
\label{wo3}
\begin{center}
\squeezetable
\begin{tabular}{llllllccccccccccc}
\hline
\hline
Phase &&&&&  $a$ (\r{A}) & $b$ (\r{A})& $c$ (\r{A}) &&& E$_g$ (eV) &P$_s$ ($\mu$C$\cdot$cm$^{-2}$)\\
\hline
 $Pm\bar{3}m$&&&&& 3.81&3.81&3.81&&&0.52 & 0\\
 $P4mm$&&&&& 3.79&3.79 &3.89&&&0.54 & 55.10 \\ 
 $Amm2$&&&&& 5.47&5.45&3.77&&&1.25 & 65.71\\  
 \hline
\hline
\end{tabular}
\end{center}
\end{table*}

\subsection{B. The origin of Rashba splitting}

In this section, combining minimal and extended models with first-principles calculations, we shed light on the origin of Rashba splitting in simple ferroelectric perovskites.
 
 \subsubsection{1. Minimal model}
 
Since the lowest conduction bands of the cubic (undistorted) phase consist in triply degenerate $t_{2g}$ states, which are split from higher-energy doubly degenerate $e_g$ states due to the octahedral crystal field $\Delta_o$, we will consider at first an effective model for $t_{2g}$=$\{yz,zx,xy\}$ electrons only. In the high-symmetry phase, hopping to neighboring transition-metal ions is mediated by bridging oxygen $p$ states, being strongly direction-dependent and resulting in substantially decoupled bonding networks for the three $t_{2g}$ bands. Using Slater-Koster parametrization to keep track of the angular dependence of hopping interactions\cite{SlaterKoster1954,Khalsa2013,Shanavas2014}, the unperturbed Hamiltonian $\mathcal{H}_0$  is diagonal in the $t_{2g}$ manifold with eigenvalues:
\begin{eqnarray}\label{eq:bare}
\varepsilon_{\alpha\beta} &=& -2t_0\left(\cos k_\alpha+\cos k_\beta \right)
\end{eqnarray}
where $\alpha,\beta=x,y,z$ and $t_0$=$t^2_{pd}/\Delta_{pd}$ is taken as the energy reference, being $t_{pd}$ and $\Delta_{pd}$ the hopping amplitude and the splitting between O-$p$ and metal-$d$ orbital states, respectively.

\begin{figure}[htb]
 \centering
 \includegraphics[width=10.0cm,keepaspectratio=true]{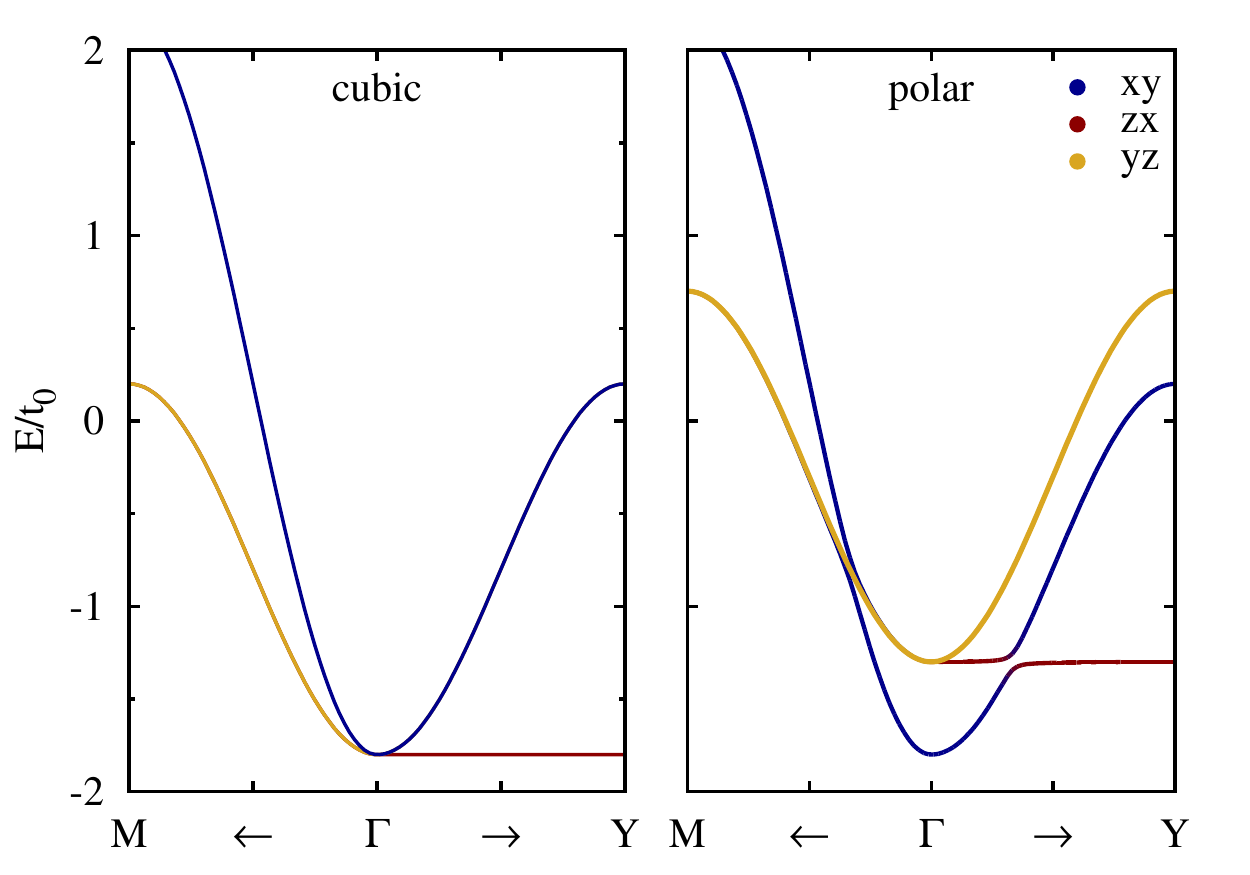}
 \caption{(Color online) In-plane band-structure as obtained with the minimal model including Eqs. (\ref{eq:bare}), (\ref{eq:fe.ham}) with $\gamma_3=0.05 t_0$ and $\Delta_{FE}=0.5t_0$, relevant for the undistorted cubic (left) and distorted tetragonal polar (right) phases, respectively. }
 \label{fig:modelbands}
\end{figure}

A polar distortion along, say, the $z$ direction has two major effects on the band structure of a cubic perovskite: first, it lifts the degeneracy within the $t_{2g}$ manifold, inducing a splitting $\Delta_{FE}$ between a lower energy $b_2=xy$ state and higher-energy doubly degenerate $e=\{yz,zx\}$ states\cite{Wolfram2006,Bersuker2006}; second, it opens new covalency channels in the metal-oxygen network due to orbital/lattice polarization effects, i.e., a polarization of the atomic-like orbital states and a change of the metal-oxygen bonding angle affecting the angular dependence of the two-center hopping integrals\cite{Khalsa2013,Shanavas2014}. Focusing only on the band-structure properties in the plane perpendicular to the polar axis, the effect of the polar distortion can be modeled by the following perturbative term:
\begin{eqnarray}\label{eq:fe.ham}
\mathcal{H}_{FE}&=&\left(\begin{array}{ccc}
\Delta_{FE} & 0 & -2 i \gamma_3 \sin k_x\\
0 & \Delta_{FE} & -2 i \gamma_3 \sin k_y\\
2 i \gamma_3 \sin k_x &2 i \gamma_3 \sin k_y &  0
\end{array}
\right)
\end{eqnarray}
where we followed the notation used in Ref. \cite{Shanavas2014} for the orbital/lattice polarization coupling term $\gamma_3$. The model accounts qualitatively very well for the effect of polar distortions onto the band structure of perovskite oxides, as shown in FIG. S \ref{fig:modelbands}.

\begin{figure}[htb]
 \centering
 \includegraphics[height=6cm]{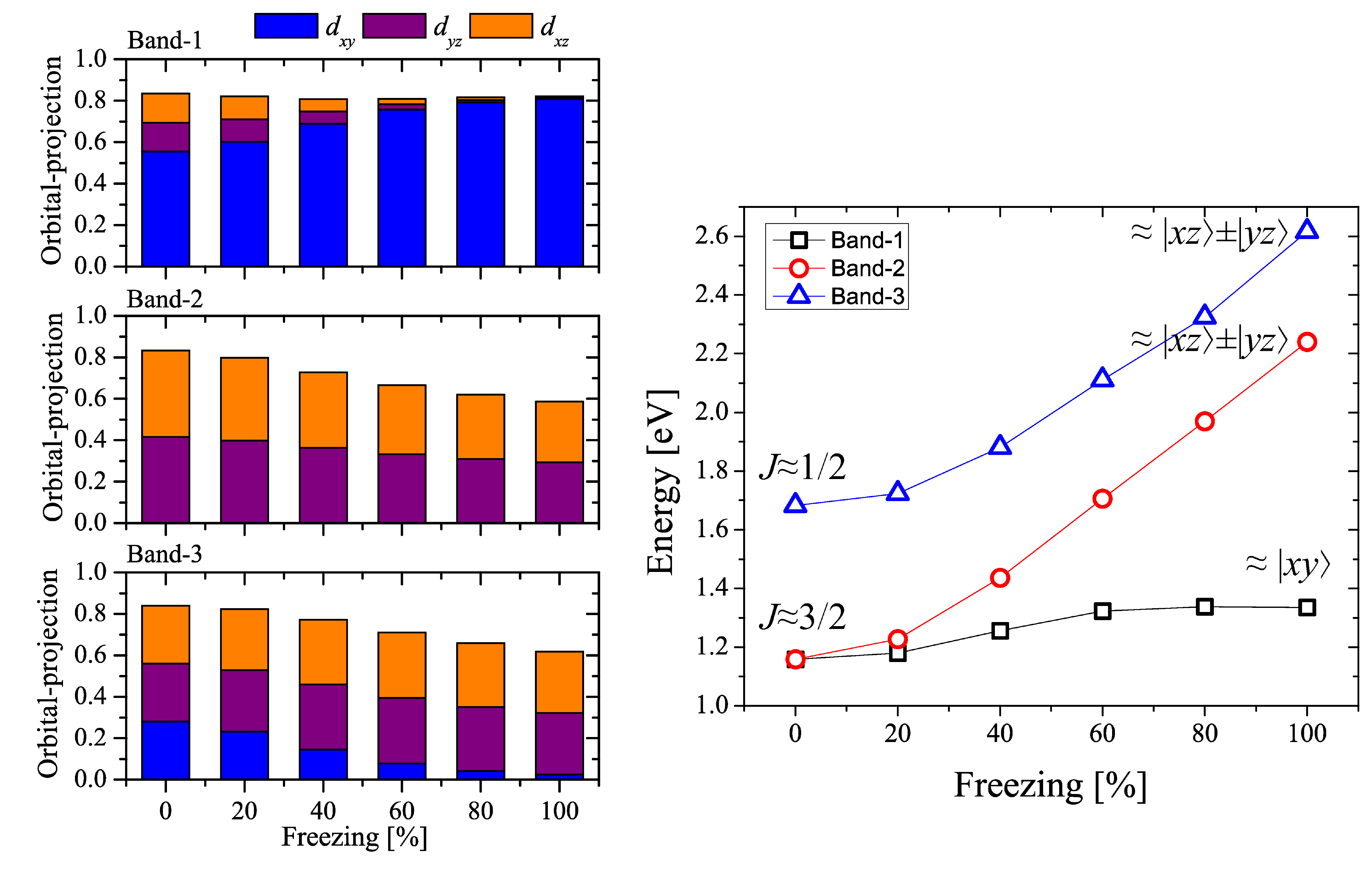}
 \includegraphics[height=6cm,keepaspectratio=true]{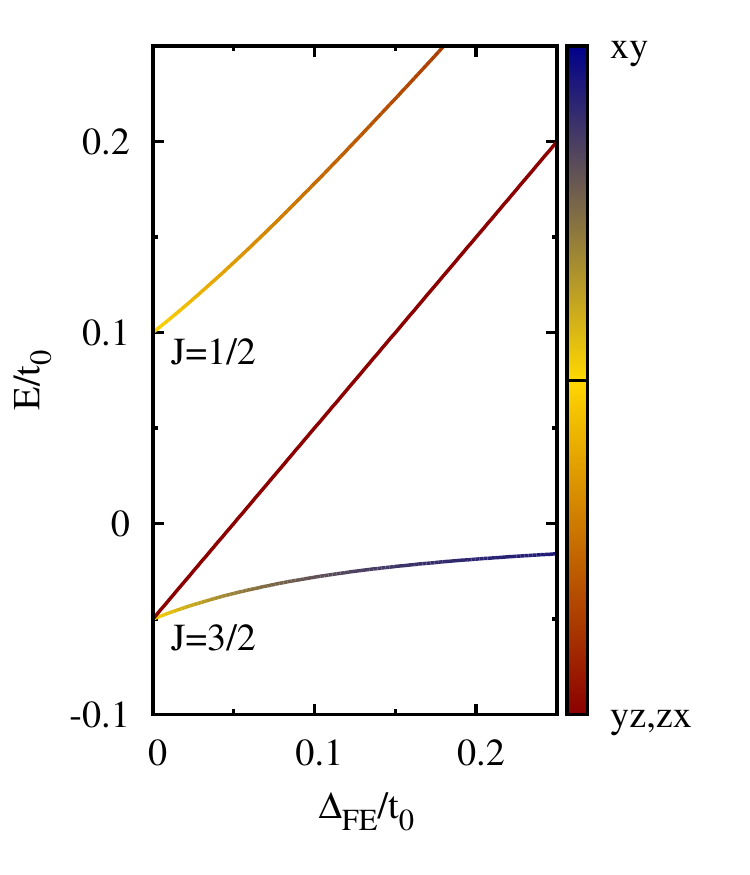}
 \caption{(Color online) Evolution of the DFT relativistic bands at $\Gamma$ point as a function of the polar distortion (left
panel) compared with the evolution of exact eigenvalues of the simplified model as a function of the CF parameter $\Delta_{FE}$  for $\lambda/t_0$ = 0.05. }
 \label{fig:modellocal}
\end{figure}

The polar-activated new hybridization channels are responsible of spin-splitting effects once spin-orbit coupling (SOC) is included. The atomic-like spin-orbit coupling for the $t_{2g}$ manifold in the $\{d_{yz},d_{zx},d_{xy}\}\otimes(\uparrow,\downarrow)$ basis reads:
\begin{eqnarray}\label{ham:soc_p}
\mathcal{H}_{soc} &=& -\lambda\left(
\begin{array}{ccc|ccc}
0 &-i & 0  & 0 & 0 & 1 \\
i & 0 & 0  & 0 & 0 & -i  \\
0 & 0 & 0  &-1 & i &  0  \\
\hline
0 & 0 & -1 & 0 & i & 0   \\
0 & 0 & -i &-i & 0 & 0   \\
1 & i &  0 & 0 & 0 & 0   \\
\end{array}\right).
\end{eqnarray}
where $\lambda$ is the SOC coupling constant. The effect of such atomic-like interaction is to split the degenerate $t_{2g}$ bands in two-fold degenerate states with total momentum $j=1/2$ and local energy $2\lambda$ and in four-fold degenerate states with total momentum $j=3/2$ and energy $-\lambda$, producing a splitting $\Delta_{SOC}=3\lambda$.
The band structure around $\Gamma$ point can be analyzed by diagonalizing the full Hamiltonian $\mathcal{H}=\mathcal{H}_0+\mathcal{H}_{FE}+\mathcal{H}_{soc}$ at $\bm k=0$ and including linear terms in $\bm k$ as subdominant contributions. The eigenvalues and eigenvectors of $\mathcal{H}$ at $\Gamma$ point are given by:
\begin{eqnarray}\label{eq:localstates}
E_1 &=& \frac{1}{2}\left(\lambda+\Delta_{FE}-\sqrt{9\lambda^2+2\lambda\Delta_{FE}+\Delta_{FE}^2} \right)\nonumber\\
E_2 &=& -\lambda+\Delta_{FE}\nonumber\\
E_3 &=& \frac{1}{2}\left(\lambda+\Delta_{FE}+\sqrt{9\lambda^2+2\lambda\Delta_{FE}+\Delta_{FE}^2} \right)
\end{eqnarray}
and
\begin{eqnarray}\label{eq:eigenstates}
\Ket{\Psi_{1s}} &=& -\sin \frac{\theta}{2}\Ket{\frac{1}{2},s\frac{1}{2}} + \cos\frac{\theta}{2}\Ket{\frac{3}{2},s\frac{1}{2}}\nonumber\\
\Ket{\Psi_{2s}} &=& \Ket{ \frac{3}{2},s\frac{3}{2}}\nonumber\\
\Ket{\Psi_{3s}} &=& \cos \frac{\theta}{2}\Ket{\frac{1}{2},s\frac{1}{2}} + \sin\frac{\theta}{2}\Ket{\frac{3}{2},s\frac{1}{2}}
\end{eqnarray}
where $s=\pm$ and $\tan\theta=2\sqrt{2}\Delta_{FE}/(\Delta_{FE}+9\lambda$), while the eigenvectors are expressed in the basis of $\vert \bm j,j_z\rangle$ states:
\begin{eqnarray}\label{eq:jj_basis}
\Ket{ \frac{1}{2},s\frac{1}{2}} &=& \frac{1}{\sqrt{3}}\left( \Ket{\psi_{yz,\bar{s}}} +  i\,s\,\Ket{\psi_{zx,\bar{s}}} + s \Ket{\psi_{xy,s}}  \right)\nonumber\\
\Ket{ \frac{3}{2},s\frac{1}{2} } &=& \frac{1}{\sqrt{6}}\left(  \Ket{\psi_{yz,\bar{s}}} + i\,s\,\Ket{\psi_{zx,\bar{s}}} - 2 s \,\Ket{\psi_{xy,s}} \right)\nonumber\\
\Ket{ \frac{3}{2},s\frac{3}{2} } &=& \frac{1}{\sqrt{2}}\left( \Ket{\psi_{yz,s}} + i\,s\,\Ket{\psi_{zx,s}} \right).
\end{eqnarray}
Clearly,  $\Delta_{FE}$ couples only states with $j_z=\pm 1/2$, since the $\Ket{ \frac{3}{2},s\frac{3}{2} }$ state comprises only $yz,zx$ orbital states. In the limit of vanishing $\Delta_{FE}$, one recovers the SOC-split states, where $E_1=E_2\mapsto -\lambda$ belonging to the $j=3/2$ manifold while $E_3\mapsto 2\lambda$. On the other hand, the ferroelectric crystal field $\Delta_{FE}$ affects the SOC-induced mixing of the $t_{2g}$ states, reducing the mixed character of the relativistic eigenstates between $\vert \psi_{xy,s}\rangle$ and $\vert \psi_{yz(zx),\bar{s}}\rangle$ which is linked to the Rashba-like spin-splitting effects.
The $xy$ character of $\Ket{\Psi_{1s}}$ and $\Ket{\Psi_{3s}}$ as a function of $\Delta_{FE}$ can be exactly evaluated, being:
\begin{eqnarray}
\vert\langle d_{xy} \vert \Psi_1\rangle\vert^2 &=&\frac{1}{2}\left(1+\frac{\lambda+\Delta_{FE}}{\sqrt{9\lambda^2+2\lambda\Delta_{FE}+\Delta_{FE}^2}} \right)\simeq  1-\frac{2}{9}\left(\frac{\Delta_{SOC}}{\Delta_{FE}}\right)^2\nonumber\\
\vert\langle d_{xy} \vert \Psi_3\rangle\vert^2 &=&\frac{1}{2}\left(1-\frac{\lambda+\Delta_{FE}}{\sqrt{9\lambda^2+2\lambda\Delta_{FE}+\Delta_{FE}^2}} \right)\simeq  \frac{2}{9}\left(\frac{\Delta_{SOC}}{\Delta_{FE}}\right)^2
\end{eqnarray}
highlighting the fact that the lowest (highest) band acquires a rapidly increasing (decreasing) pure $d_{xy}$ character, as also shown in FIG. S \ref{fig:modellocal}.

In the rotated basis the $\bm k$ dependence of the Hamiltonian up to linear order in crystal momentum reads:
\begin{eqnarray}\label{ham:kdep}
\mathcal{H}_\Gamma &=&
\left(\begin{array}{ccc}
E_1\sigma_0+\alpha_{11}(\sigma_x\,k_y-\sigma_y\,k_x) &\alpha_{12}(\sigma_z\,k_y+i\sigma_0\,k_x) &\alpha_{13}(\sigma_x\,k_y-\sigma_y\,k_x)\\
\alpha_{12}(\sigma_z\,k_y-i\sigma_0\,k_x)&E_2\sigma_0&\alpha_{23}(\sigma_z\,k_y-i\sigma_0\,k_x)\\
\alpha_{13}(\sigma_x\,k_y-\sigma_y\,k_x)&\alpha_{23}(\sigma_z\,k_y+i\sigma_0\,k_x)& E_3\sigma_0+\alpha_{33}(\sigma_x\,k_y-\sigma_y\,k_x)
\end{array}\right)
\end{eqnarray}
where $\bm \sigma$ are Pauli matrices and $\sigma_0$ is the 2$\times$2 identity matrix. Since the three manifolds are well separated in energy, and all off-diagonal terms are already linear in $k$, the leading term of the spin-momentum coupling is that parametrized by the diagonal Rashba parameters $\alpha_{ii}$, while the effect of the off-diagonal terms could be included using the standard L\"owdin partitioning\cite{lowdin1951}, resulting in cubic spin-momentum coupling terms. The Rashba coefficient for the lowest-energy state is given by:
\begin{eqnarray}\label{eq:1rashba}
\alpha_{11} &=& \gamma_3\frac{2}{3}\left(\frac{1}{\sqrt{2}}\sin \theta-2\cos\theta\right)\equiv-12\,\gamma_3\,\frac{\lambda}{\sqrt{9\lambda^2+2\lambda\Delta_{FE}+\Delta_{FE}^2}}  \simeq-4\,\gamma_3\,\frac{\Delta_{SOC}}{\Delta_{FE}}
\end{eqnarray}
where the last result is valid in the limit of $\Delta_{SOC}/\Delta_{FE}\ll 1$. Therefore, the Rashba coupling of the lowest-energy state is controlled by the same parameter $\Delta_{SOC}/\Delta_{FE}$ which measures its pure $xy$ character. In the limit of very large $\Delta_{FE}\gg\Delta_{SOC}$, the Rashba splitting vanishes, simply because there is no effective coupling between opposite-spin manifolds.

On the one hand, the  minimal model considered so far allows to explain why the lowest-energy state shows a negligibly small spin-splitting in the presence of a sizeable polar distortion. In fact, the Rashba coupling constant appears to be directly proportional to the activated new covalency channel (modeled by the $\gamma_3$ parameter), at the same time being inversely proportional to the energy splitting $\Delta_{FE}$; the competing and compensating effects of these two physical ingredients lead to a subtstantial suppression of spin splitting in the $E_1$  manifold. On the other hand, it fails in reproducing the large spin splittings observed in the $E_2, E_3$ manifolds. In fact, no linear spin-momentum coupling appears in the $E_2$ manifold, while $\alpha_{33}=-\alpha_{11}$ for the $E_3$ manifold, implying its strong suppression as a function of the polar crystal field. 

 \subsubsection{2. Extended model}

As previously noticed in Ref. \cite{Held2015}, the main reason for this is that the atomic $t_{2g}$ SOC is not an accurate enough description, and $k$-dependent corrections to the purely atomic $\mathcal{H}_{soc}$ need to be included. Two such corrections, both activated by orbital/lattice polarization effects, can be envisaged: i) including the effect of SOC onto bridging oxygen ions mediating the $d$-$d$ hopping interactions or ii) including contributions from $e_g$ states. The former results in spin-flip hopping terms between $yz/zx \uparrow$ and $yz/zx \downarrow$ states\cite{Held2015}, giving
\begin{eqnarray}\label{ham:soc_oxy}
\mathcal{H}_{soc}^{(O)} &=& 2 \gamma^{(0)}\left(
\begin{array}{ccc|ccc}
0&0&0&\sin k_y & 0&0\\
0&0&0&0&i\sin k_x & 0 \\
0&0&0&0&0&0\\
\sin k_y&0&0&0&0&0\\
0&-i\sin k_x&0&0&0&0\\
0&0 & 0&0&0&0
\end{array}
\right)
\end{eqnarray}
Such Hamiltonian provides the following corrections to the diagonal Rashba coupling terms 
\begin{eqnarray}
\alpha^{(O)}_{11} &=& \gamma^{(O)}\frac{1}{3}\left( 2\sqrt{2}\sin\theta+\cos\theta- 3\right) = \gamma^{(O)}\left(\frac{\Delta_{FE}+\lambda}{\sqrt{9\lambda^2+2\lambda\Delta_{FE}+\Delta_{FE}^2}}-1 \right) \nonumber\\ &\simeq& -\frac{4}{9}\gamma^{(O)}\left(\frac{\Delta_{SOC}}{\Delta_{FE}}\right)^2\nonumber\\
\alpha_{22}^{(O)}&=&\gamma^{(O)} \nonumber\\
\alpha^{(O)}_{33} &=& \gamma^{(O)}\frac{1}{3}\left( 2\sqrt{2}\sin\theta+\cos\theta + 3\right) = \gamma^{(O)}\left(\frac{\Delta_{FE}+\lambda}{\sqrt{9\lambda^2+2\lambda\Delta_{FE}+\Delta_{FE}^2}}+1 \right) \nonumber\\ &\simeq& \gamma^{(O)}\left(\frac{2}{3}-\frac{4}{9}\left(\frac{\Delta_{SOC}}{\Delta_{FE}}\right)^2\right)\nonumber\\
\end{eqnarray}
A Rashba-like spin-momentum coupling emerges in the $E_2$ manifold with a coupling constant which is directly proportional to the inversion asymmetry correction of the O-mediated hopping interactions. It is worth to notice that while the Rashba coupling in the lowest-energy state is still found to vanish as $\Delta_{SOC}/\Delta_{FE}\ll 1$, a finite coupling term survives in the $E_3$ manifold.

When considering the virtual processes between $t_{2g}$ and $e_g$ manifolds, the additional hopping interactions $\gamma_1$ and $\gamma_2$, arising from orbital/lattice polarization effects between the $\{yz/zx\}$ and $z^2$ and $x^2-y^2$ orbital states, respectively\cite{Shanavas2014}, must be included alongside the atomic SOC interacting terms; the resulting $k-$dependent SOC Hamiltonian reads:
\begin{eqnarray}\label{ham:soc_eg}
\mathcal{H}_{soc}^{(e_g)} &=& \frac{2\lambda}{\Delta_o}\left(
\begin{array}{cc}
H_{\uparrow\uparrow} & H_{\uparrow\downarrow}\\
H_{\uparrow\downarrow}^\dagger & H_{\downarrow\downarrow}
\end{array}
\right)
\end{eqnarray}
where
\begin{eqnarray}
H_{\uparrow\uparrow} &=& \left(
\begin{array}{ccc}
2\lambda & -i\,\lambda & 2\gamma_2\sin k_y\\
i\,\lambda & 2\lambda &-2\gamma_2\sin k_x \\
 2\gamma_2 \sin k_y &-2\gamma_2\sin k_x& 2\lambda
\end{array}\right)
\end{eqnarray}
\begin{eqnarray}
H_{\uparrow\downarrow} &=& \left(
\begin{array}{ccc}
2\left(\sqrt{3}\gamma_1-\gamma_2\right)\sin k_y & \left(\sqrt{3}\gamma_1+\gamma_2\right)(\sin k_x+i\,\sin k_y) &\lambda\\
\left(\sqrt{3}\gamma_1+\gamma_2\right)(\sin k_x+i\,\sin k_y) & 2 i \left(\sqrt{3}\gamma_1-\gamma_2\right)\sin k_x& -i\,\lambda\\
-\lambda & i\,\lambda &0
\end{array}\right)
\end{eqnarray}
\begin{eqnarray}
H_{\downarrow\downarrow} &=& \left(
\begin{array}{ccc}
2\lambda & i\,\lambda & -2\gamma_2\sin k_y\\
-i\,\lambda & 2\lambda &2\gamma_2\sin k_x\\
-2\gamma_2\sin k_y&2\gamma_2\sin k_x& 2\lambda
\end{array}\right)
\end{eqnarray}
This interacting term also produces Rashba-like spin-momentum coupling in all three manifolds, whose coupling constants can be expressed as:
\begin{eqnarray}
\alpha_{11}^{(eg)} &=& 2\gamma_2\frac{\lambda}{\Delta_o}(\sqrt{2}\sin\theta-\cos\theta-1) =
 2\gamma_2\frac{\lambda}{\Delta_o} \left(\frac{\Delta_{FE}-3\lambda}{\sqrt{9\lambda^2+2\lambda\Delta_{FE}+\Delta_{FE}^2}}-1\right) \nonumber\\ &\simeq& \frac{8}{9}\,\gamma_2\,\frac{\Delta_{SOC}^2}{\Delta_o\Delta_{FE}}\nonumber\\
\alpha_{22}^{(eg)} &=& \frac{4}{\sqrt{3}}\,\gamma_1\,\frac{\Delta_{SOC}}{\Delta_o}\nonumber\\
\alpha_{33}^{(eg)} &=& 2\gamma_2\frac{\lambda}{\Delta_o}(\sqrt{2}\sin\theta-\cos\theta+1) =  2\gamma_2\frac{\lambda}{\Delta_o} \left(\frac{\Delta_{FE}-3\lambda}{\sqrt{9\lambda^2+2\lambda\Delta_{FE}+\Delta_{FE}^2}}+1\right) \nonumber\\ &\simeq& \frac{2}{3}\,\gamma_2\,\frac{\Delta_{SOC}}{\Delta_o}\left(1-\frac{1}{3}\frac{\Delta_{SOC}}{\Delta_{FE}}\right)\nonumber\\
\label{equ_twy_use}
\end{eqnarray}
In this case, the Rashba-like splitting in the $E_2$ ($E_3$) manifold would raise mostly from virtual processes with the $z^2$ ($x^2-y^2$) orbital state, while the splitting in the lowest-energy state is again found to vanish as the crystal-field splitting $\Delta_{FE}$ increases.

Given the several energy scales entering in the proposed model, identifying the most relevant mechanism responsible for the observed spin splitting is not a trivial task. It is worth to notice, however, that the spin-splitting of the band connected to the lowest-energy state is always found to be strongly suppressed as its character becomes strongly $xy$-type, i.e., perpendicular to the polar axis. 

 \subsubsection{3. First-principles results}

\begin{figure}[h!]
 \centering
 \includegraphics[height=18cm]{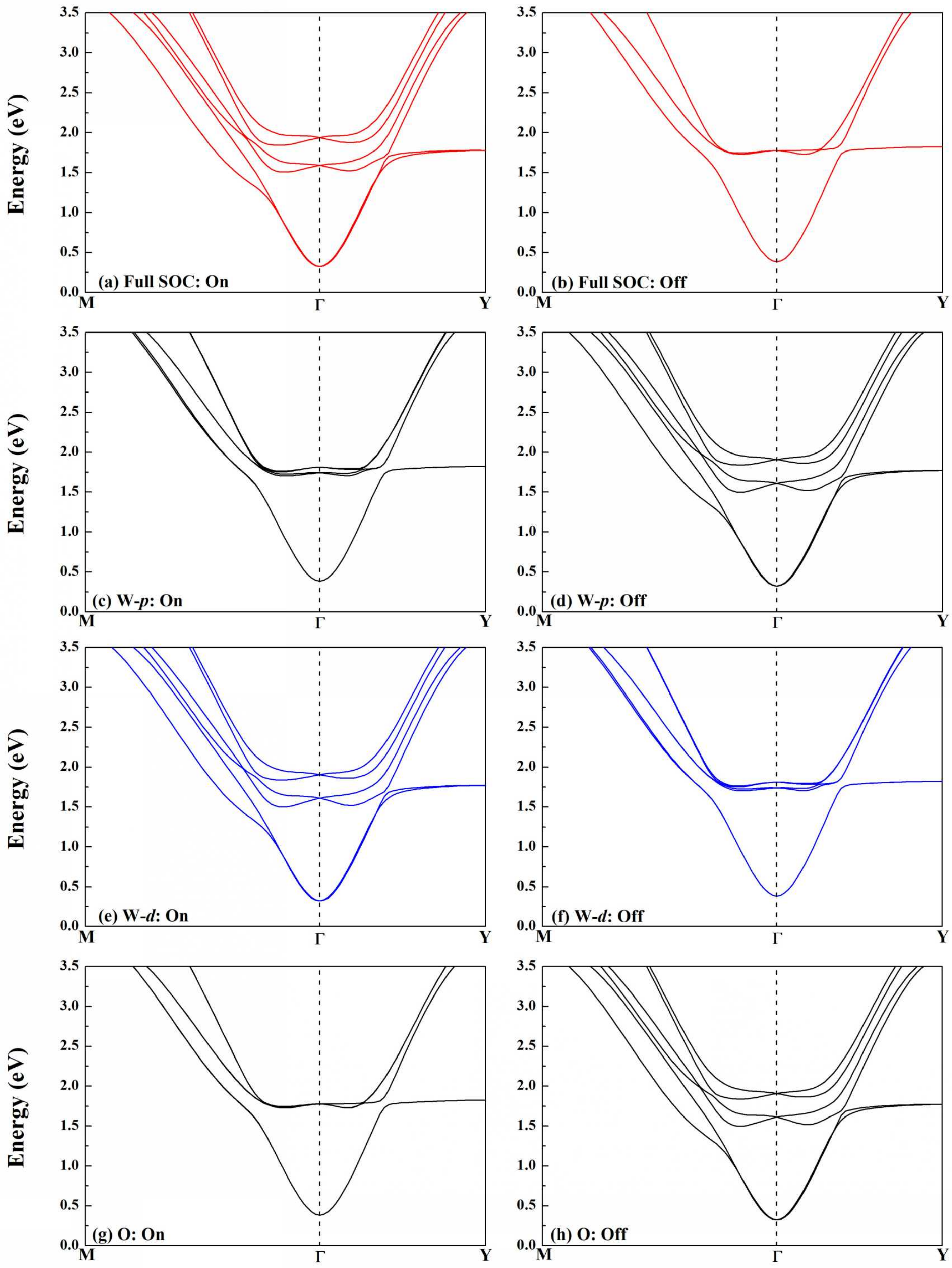}
 \caption{(Color online) The band structures for the $P4mm$ phase of WO$_3$ (a) with full SOC, (b) without SOC,  (c) with and (d) without SOC from W-$p$ orbitals, (e) with and (f) without SOC from W-$d$ orbitals, (g) with and (d) without SOC from oxygens.}
 \label{fig:vaspsoc}
\end{figure}

In order to analyse the origin of Rashba splitting more quantatively, we then carry out first-principles calculations. First of all, we artificially switch on/off the partial SOC matrix elements to directly invistigate the orbital dependence of the Rashba splitting in the $P4mm$ phase of WO$_3$. When we compare in FIG. S \ref{fig:vaspsoc} panel (a) with (h) and panel (b) with (g), we observe that they are exactly the same, indicating that oxygens produce no significant Rashba splitting. Although the band structures without SOC from W-$p$ orbitals (FIG. S \ref{fig:vaspsoc}(d)) look quite similar to the one with full SOC (FIG. S \ref{fig:vaspsoc}(a)), the little difference between FIG. S \ref{fig:vaspsoc}(c) and (b) demonstrates the existence of minor contributions from W-$p$ states. If we just consider the SOC matrix elements from W-$d$ orbitals, there already exists obvious Rashba splitting in FIG. S \ref{fig:vaspsoc}(e). Reversely, for the case without SOC from W-$d$ orbitals (FIG. S \ref{fig:vaspsoc}(f)), the Rashba-type splitting is drastically suppressed. It is interesting to point out that FIG. S \ref{fig:vaspsoc}(f) and (c) share almost the same characteristics, which proves the effective influence of W-$p$ orbitals on Rashba splitting. Surely, it is tiny and negligible, in contrast to that from W-$d$ states. In a word, we can conclude from the FIG. S \ref{fig:vaspsoc} that $d$ orbitals of tungstens play the dominant effects in the large Rashba splitting here.

\begin{figure}[htb]
 \centering
 \includegraphics[height=7.5cm]{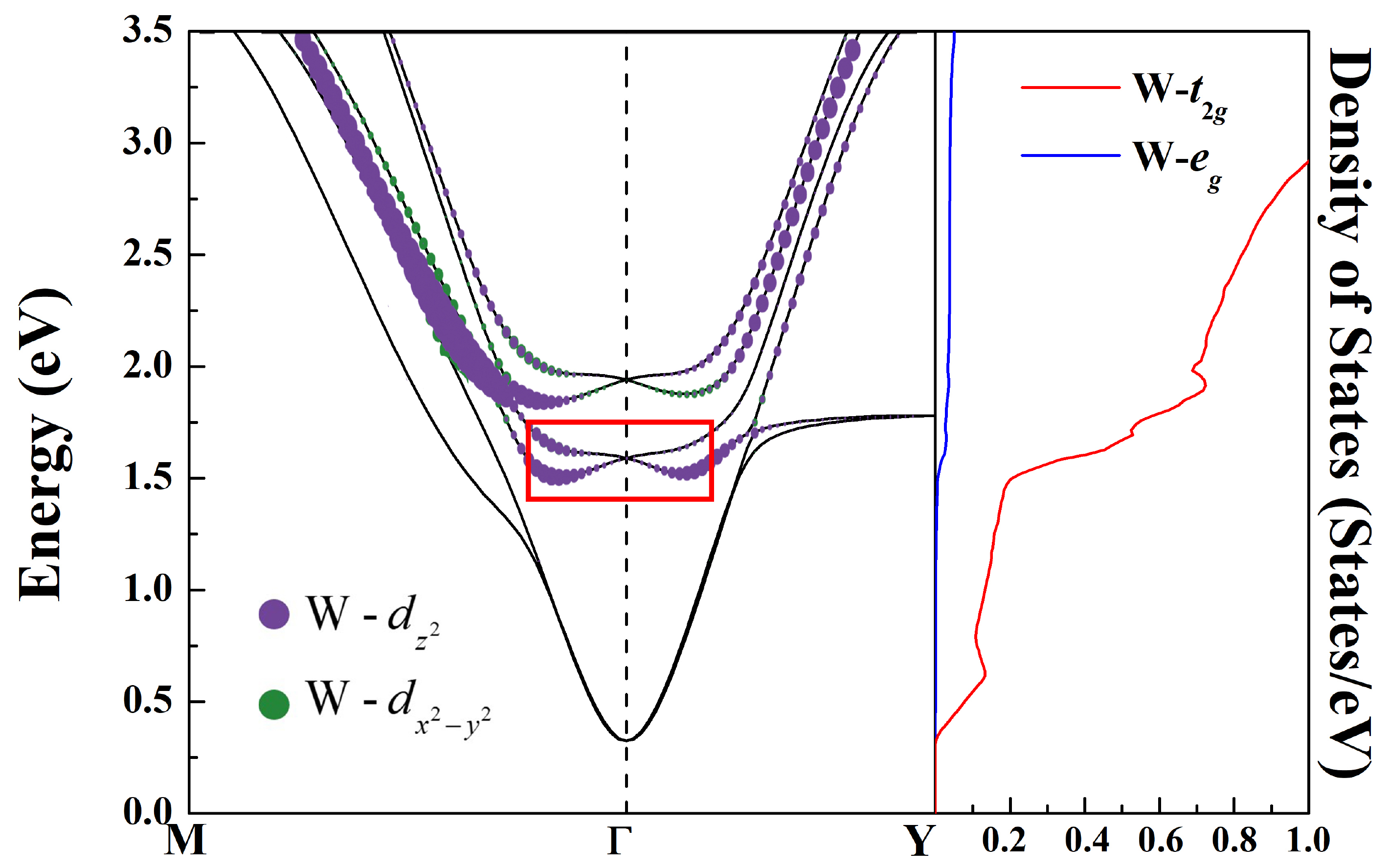}
 \caption{(Color online) Left panel is the orbital projected band structures for the $P4mm$ phase of WO$_3$ with W-$d_{z^2}$ orbital in purple and W-$d_{x^2-y^2}$ orbital in green. The radius of the circles indicates the weight of the orbitals. Right panel is the corresponding density of states.}
 \label{fig:vaspdos}
\end{figure}

According to our previous analytical model, such an effect can be closely related to the hybridization between $t_{2g}$ and $e_{g}$ states, which is confirmed by our density-of-states calculations. As clearly shown in FIG. S \ref{fig:vaspdos}, for the lowest $E_1$ band, there is no W-$e_g$ components, resulting in the absence of Rashba splitting. However, the minor occupied $e_g$ states hybridize with the dominant $t_{2g}$ orbitals for the higher $E_2$ and $E_3$ bands, and then trigger their large Rashba-type splitting. Especially, the hybridization for the $E_2$ band is mainly between $t_{2g}$ and W--$d_{z^2}$ states, in accordance to the EQ(\ref{equ_twy_use}).

\begin{figure}[htb]
 \centering
 \includegraphics[height=12.5cm]{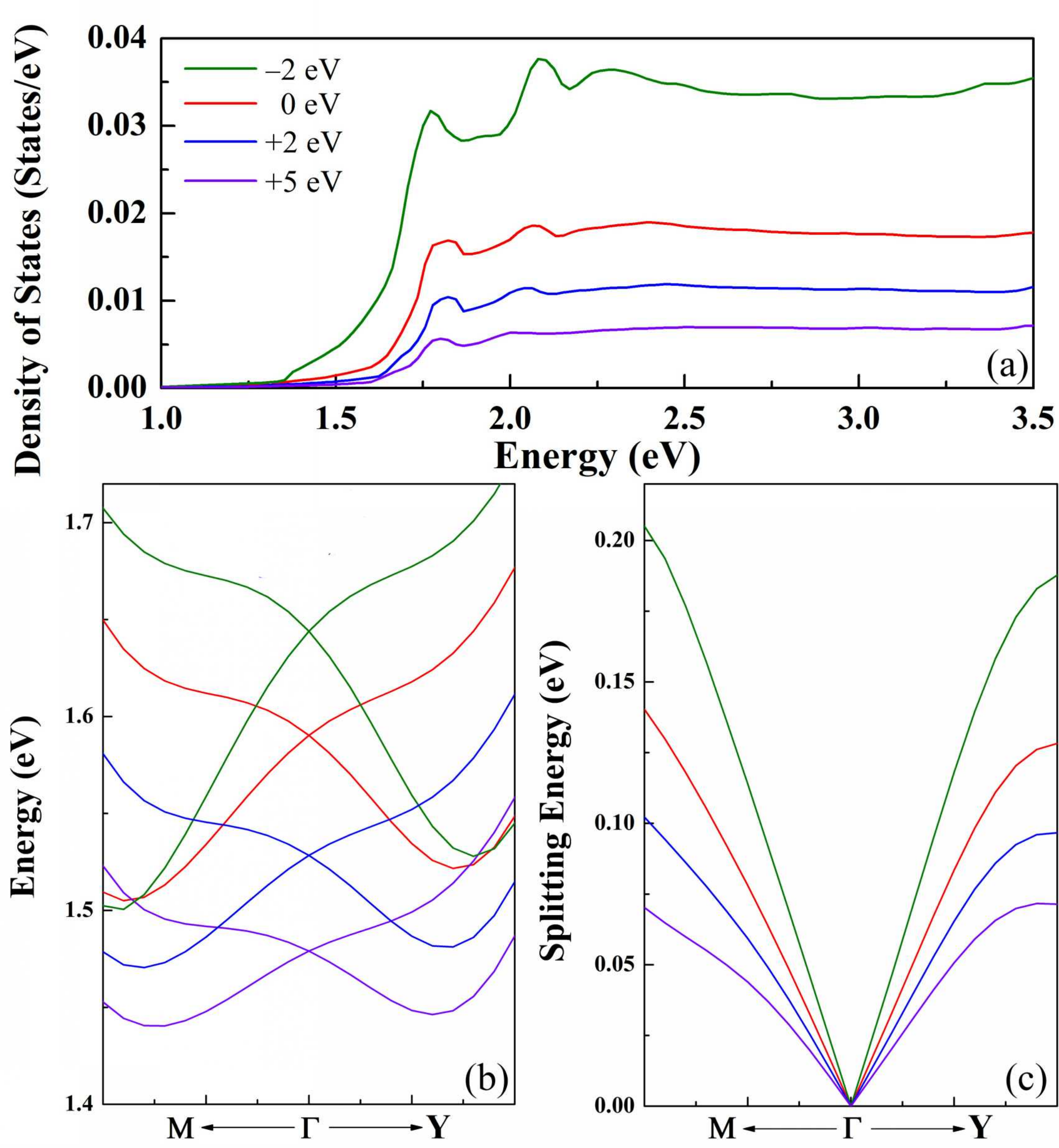}
 \caption{(Color online) With the externally applied orbital selective potential energy $V_{\rm ext}^{d_{z^2}}$ = 0, 5 and ${\pm 2}$ eV, (a) density of states for W-$d_{z^2}$ states, (b) electronic dispersion curves, and (c) the corresponding Rashba-typle splitting energy for the $E_2$ band in the $P4mm$ phase of WO$_3$. The bands in red of (c) correspond to those in red square of FIG. S \ref{fig:vaspdos}(a). The Fermi level is set to zero.}
 \label{fig:vasplocal}
\end{figure}

To directly assess the influence of $e_{g}$ states on Rashba splitting, we then use orbital selective external potential (OSEP) method\cite{Xiangang,Yongping}. This approach can introduce a special external potential on any selected orbitals, which is analogous to the DFT + $U$ method\cite{Vladimir}.Within the frame of OSEP, the specifically assigned atomic orbit $|inlm\sigma\rangle$ feels the potential $V_{\rm ext}$, making the system Hamiltonian become as follows:
\begin{eqnarray}\label{eq:bare}
H^{\rm OSEP} &=& H_{\rm KS}^0 +  {|inlm\sigma\rangle}{\langle}{inlm\sigma|}  V_{\rm ext}
\end{eqnarray}
where $i$ denotes the atomic site, and $n,l,m,\sigma$ are the principle, orbital , magnetic and spin quantum numbers, respectively. $ H_{\rm KS}^0$ is the unperturbed Kohn-Sham Hamiltonian. Since the strength of overlap between orbitals is strongly dependent on their energy difference, we can modify the orbital interaction between states by applying an artificial field to shift the energy levels. As a representative, in FIG. S \ref{fig:vasplocal}, we shift the energy level of the W-$d_{z^2}$ orbital in the $P4mm$ phase of WO$_3$ to investigate its influence on Rashba-like splitting of the $E_2$ band.

It is clealry shown in  FIG. S \ref{fig:vasplocal}(a) that the artificial field indeed shifts the energy level of the W-$d_{z^2}$ orbital and greatly affects the DOS. When the potential energy $V_{\rm ext}^{d_{z^2}}$ changes from negative to positive, the position of $z^2$ states moves upward. Their density of states is progressively reduced during the process, indicating that the hybridization between $t_{2g}$ and $z^2$ states becomes smaller. When an artificial field is applied to up-shift $z^2$ orbital by 2 eV, due to its weaker hybridization with the $t_{2g}$ states, the Rashba-type splitting in the $E_2$ band is suppressed. Such a large external potential will also bring the Fermi level upward, therefore the $E_2$ band is equivalently shifted down as shown in  FIG. S \ref{fig:vasplocal}(b). With the reduction of the hybridization between $t_{2g}$ and $z^2$ orbitals by enlarging the positive $V_{\rm ext}$ to 5 eV, the $E_2$ band moves downward continuously. Meanwhile, its Rashba splitting further decreases. Inversely, if we introduce a negative field to the $P4mm$ phase of WO$_3$ ( $V_{\rm ext}^{d_{z^2}}$ = $-$2 eV), the energy level of $z^2$ orbital is shifted down. As expected, stronger orbital interaction between $t_{2g}$ and $z^2$ states results in an enlarged Rashba splitting in electronic dispersion for the  $E_2$ band. In order to summarize the influence of applied potential on the Rashba-type splitting, we compare the splitting energy in  FIG. S \ref{fig:vasplocal}(c). It is clear that with the enhancement of the external field applying to the W-$d_{z^2}$ orbital, the splitting energy of the $E_2$ band is gradually decreased. We therefore can conclude that the orbital interaction between $t_{2g}$ and $z^2$ states can is the main feature responsible for the Rashba-type splitting, which is consistent with our analytical model and gives a strong evidence to confirm the critical role of $e_g$ states on Rashba splitting.

\subsection{C. Ferroelectric and Rashba splittings in the $Amm2$ phase}

\begin{figure}[htb]
 \centering
 \includegraphics[width=15.0cm,keepaspectratio=true]{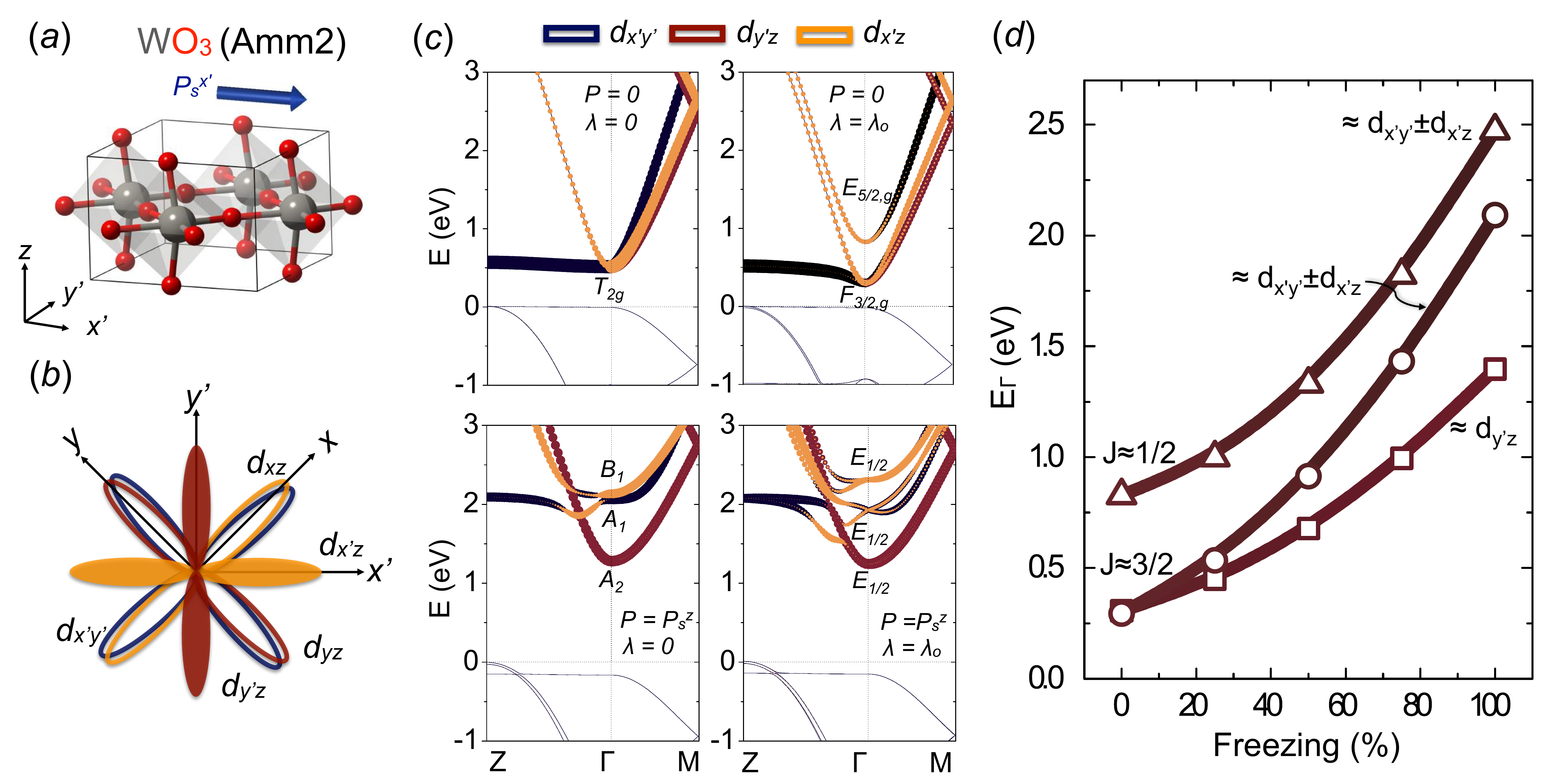}
 \caption{(Color online) ($a$) Sketch of the atomic structure of the $Amm2$ phase of WO$_3$. ($b$) Rotated reference with the polarization oriented along the $x'$ direction. ($c$) Electronic bands for the $Pm\bar{3}m$ and $Amm2$ phases of WO$_3$ with and without SOC. ($d$) Evolution of the lowest conduction state at $\Gamma$  in terms of the amplitude of the ``freezed'' polar distortion into $Pm\bar{3}m$, when including SOC (as explained in FIG.1b). }
 \label{S1}
\end{figure}

In FIG. S \ref{S1}, we describe the evolution of lowest conduction states in terms of the polar distortion and SOC  in the orthorhombic $Amm2$ phase of the WO$_3$. Here, the spontaneous polarisation is along the $x'$-axis rotated 45$^o$ around the $z$-axis respect to the $x$-axis of the cubic perovskite. As observed in FIG. S \ref{S1}, in the presence of P$_s^{x'}$, the orthorhombic symmetry lift the degeneracy of the $t_{2g}$ states into three states associated to $d_{y'z}$ perpendicular to the polarization, $d_{x'y'}$ and $d_{x'z}$. Similarly to $P4mm$, the $SOC$ mixes the three $t_{2g}$ orbitals. As the polarization increases, the lowest band acquires a dominant $d_{y'z}$ character, while the higher-energy levels mix $d_{x'y'}$ and $d_{x'z}$ orbitals.

\hfill


\begin{center}
\section{II. The case of KT$a$O$_3$}
\end{center}

\subsection{A. Relaxed polar structures}

KTaO$_3$ is known to be an incipient ferroelectric,  which can however be made ferroelectric under strain engineering \cite{PhysRevLett.104.227601}. Under suficcient tensile epitaxial biaxial strain, it adopts a ferroelectric $Amm2$ ground state with polarization in-plane (along the $xy$-direction) ; under sufficient compressive epitaxial biaxial strain, it adopts a ferroelectric $P4mm$ ground state with polarization out-of-plane (along the $z$-direction).  TABLE S \ref{kto} report the main characteristics of KTaO$_3$ in its reference $Pm\bar{3}m$ cubic phase and in the $P4mm$ and $Amm2$ polar phases obtained under $-3$\% and $+3$\% epitaxial biaxial strain.


\begin{table*}[h!]
\caption{Computed lattice parameters and polarization of KTaO$_3$ under different epitaxial strain conditions.} 
\label{kto}
\begin{center}
\squeezetable
\begin{tabular}{llllllccccccccccc}
\hline
\hline
Phase  &Strain (\%) &&&&  $a$ (\r{A}) & $b$ (\r{A})& $c$ (\r{A}) &&&  &P$_s$ ($\mu$C$\cdot$cm$^{-2}$)\\
\hline
 $Pm\bar{3}m$&(0 \%) &&&& 3.96 & 3.96 & 3.96 &&& & 0\\
 $P4mm$&(-3\%) &&&& 3.88 & 3.88 & 4.13  &&& & 27.07\\
 $Amm2$& (+3\%) &&& & 5.82 & 5.82 & 3.94 &&& & 38.02\\  
 \hline
\hline
\end{tabular}
\end{center}
\end{table*}

\subsection{B. Ferroelectric and Rahsba splittings in the strained $P4mm$ and $Amm2$ phases}

\begin{figure}[htb]
 \centering
 \includegraphics[width=16.0cm,keepaspectratio=true]{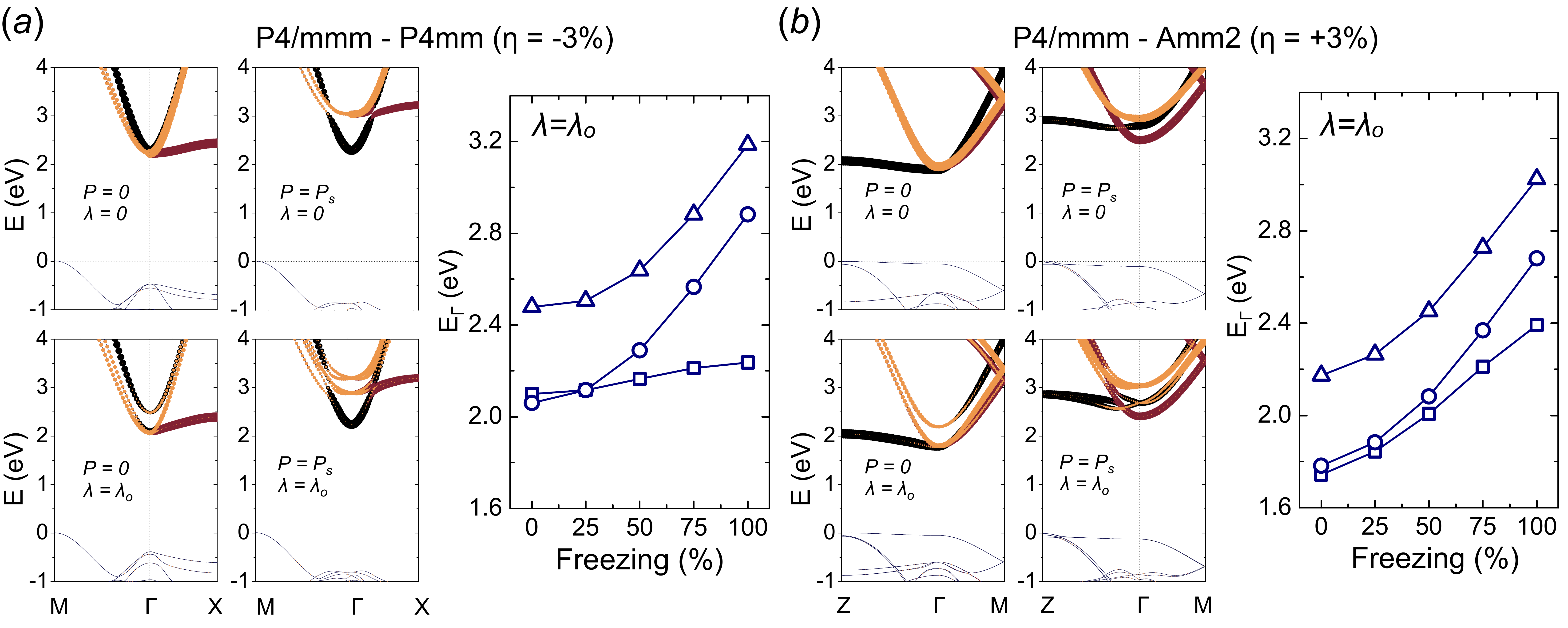}
 \caption{(Color online) Evolution of the electronic band structure of KTaO$_3$ in terms of SOC and polar distortion when going from $P4/mmm$ to $P4mm$ phase under $-3$\% strain ($a$) and from $P4/mmm$ to$Amm2$ under $+3$\% strain ($b$).}
 \label{S2}
\end{figure}

In FIG. S\ref{S2}, we report the evolution electronic band-structure of KTaO$_3$ in terms of SOC and polar distortion when going (i) from the $P4/mmm$ paraelectric reference to the $P4mm$ polar ground state under $-3$\% epitaxial biaxial strain and (ii)  from the $P4/mmm$ paraelectric reference to the $Amm2$ polar ground state under $+3$\% epitaxial biaxial strain. We identify the same behavior as for WO$_3$. Our results are in line with those reported by Tao and Wang in Ref. \onlinecite{doi:10.1063/1.4972198}.  
 

\hfill

\begin{center}
\section{III. The case of Bi$_2$WO$_6$}
\end{center}

\subsection{A. Relaxed polar structures}

In TABLE S \ref{amplimodes}, we report the cell parameters and the amplitude of the atomic distortions with respect to the paraelectric $I4/mmm$ reference (expressed in terms of symmetry-adapted mode amplitudes as obtained with AMPLIMODE  \cite{Perez-Mato:sh5107}) for the four ferroelectric metastable phases. We report amplitudes of primary distortions ($\Gamma_5^{-}$,  $X_2^{+}$ and $X_3^{+}$) as well as of secondary modes ($X_1^{-}$, $X_3^{-}$, $M_5^{+}$ and $M_3^{-}$).  The relaxed structures of the $B2cb$ and$P2_1ab$ phases are in good agreement with the experiment (disciption typically better in PBEsol than in LDA). We see no significant effect of the SOC on the atomic structure.


\begin{table*}[!h]
\begin{center}
\squeezetable
\caption{Symmetry-adapted mode amplitudes (\r{A}) respect the $I4/mmm$ paraelectric reference and lattice parameters (\r{A}) of distinct polar phases of Bi$_2$WO$_6$ fully relaxed in GGA (PBEsol) with and without SOC. Mode decomposition was performed using  the AMPLIMODE software \cite{Perez-Mato:sh5107}. Modes amplitudes are in \r{A}. Comparison with experimental data \cite{McDowell} is provided for the $B2cb$ intermediate phase and $P2_1ab$ ground state. LDA relaxation are performed with ABINIT package \cite{ABINIT}.} 
\label{amplimodes} 
\begin{tabular}{lccccccccccccccccccccccc}
\hline
\hline
         & \multicolumn{3}{c}{$F2mm$}&&\multicolumn{4}{c}{$B2cb$}&&\multicolumn{3}{c}{$B2cm$}&&\multicolumn{4}{c}{$P2_1ab$} \rule[-1ex]{0pt}{3.5ex} \\
\cline{2-4}\cline{6-9}\cline{11-13}\cline{15-18}
Modes  (\r{A}) & \scriptsize SOC &   \scriptsize noSOC &  \scriptsize LDA &&  \scriptsize SOC & \scriptsize noSOC & \scriptsize Exp. & \scriptsize LDA &&  \scriptsize SOC &  \scriptsize  noSOC & \scriptsize LDA &&  \scriptsize SOC &   \scriptsize noSOC & \scriptsize Exp. & \scriptsize LDA \\
\hline
 $\Gamma_5^{-}$   & 0.65 & 0.63 &0.39&& 0.61 & 0.60 & 0.61&0.40&& 0.53 & 0.50 &0.32&& 0.57 & 0.56& 0.55&0.34\\
 $X_2^{+}$   & - & - &-&& - & - &- &-&&0.62 & 0.63 &0.61&&0.41 & 0.40 &0.36&0.53\\
  $X_3^{+}$     & - & - &-&& 0.81 & 0.82 &0.81 &0.62&&- & - &-&&0.72 & 0.73 &0.58&0.41 \\
 $X_1^{-}$  & - & - &-&& 0.088 & 0.080 &0.087 &0.037&&- & - &-&&0.07 & 0.06&0.07&0.019  \\
 $X_3^{-}$   & - & - &-&& - & - &- &-&&- &- &-&&0.04 & 0.03&0.06&0.05 \\
 $X_4^{-}$   & - & - &-&& - & - &- &-&&0.08 & 0.09 &0.10&&- & - &-&-\\
  $M_5^{+}$   & - & - &-&& - & - &- &-&&- & - &-&&0.10 & 0.10 &0.05&0.05\\
$M_3^{-}$    & - & - &-&& - & - &- &-&&- & - &-&&0.01 & 0.01&0.01&0.006\\
\hline 
Cell  (\r{A})& & \\
\hline
$a$ &5.53&5.51&5.37&&5.49&5.48&5.53&5.38&&5.42&5.40&5.29&&5.46&5.45 &5.48&5.32\\
$b$ &5.47&5.46&5.36&&5.48&5.47&5.50&5.35&&5.40&5.38&5.29&&5.45&5.45 &5.46&5.30\\
$c$ &16.39&16.40&16.00&&16.53&16.53&16.55&16.12&&16.45&16.46&16.17&&16.53&16.53&16.48&16.17\\
\hline
\hline
\end{tabular}
\end{center}
\end{table*} 

\subsection{B. Electronic band structure }

\begin{figure}[h!tb]
 \centering
 \includegraphics [width=12cm,keepaspectratio=true]{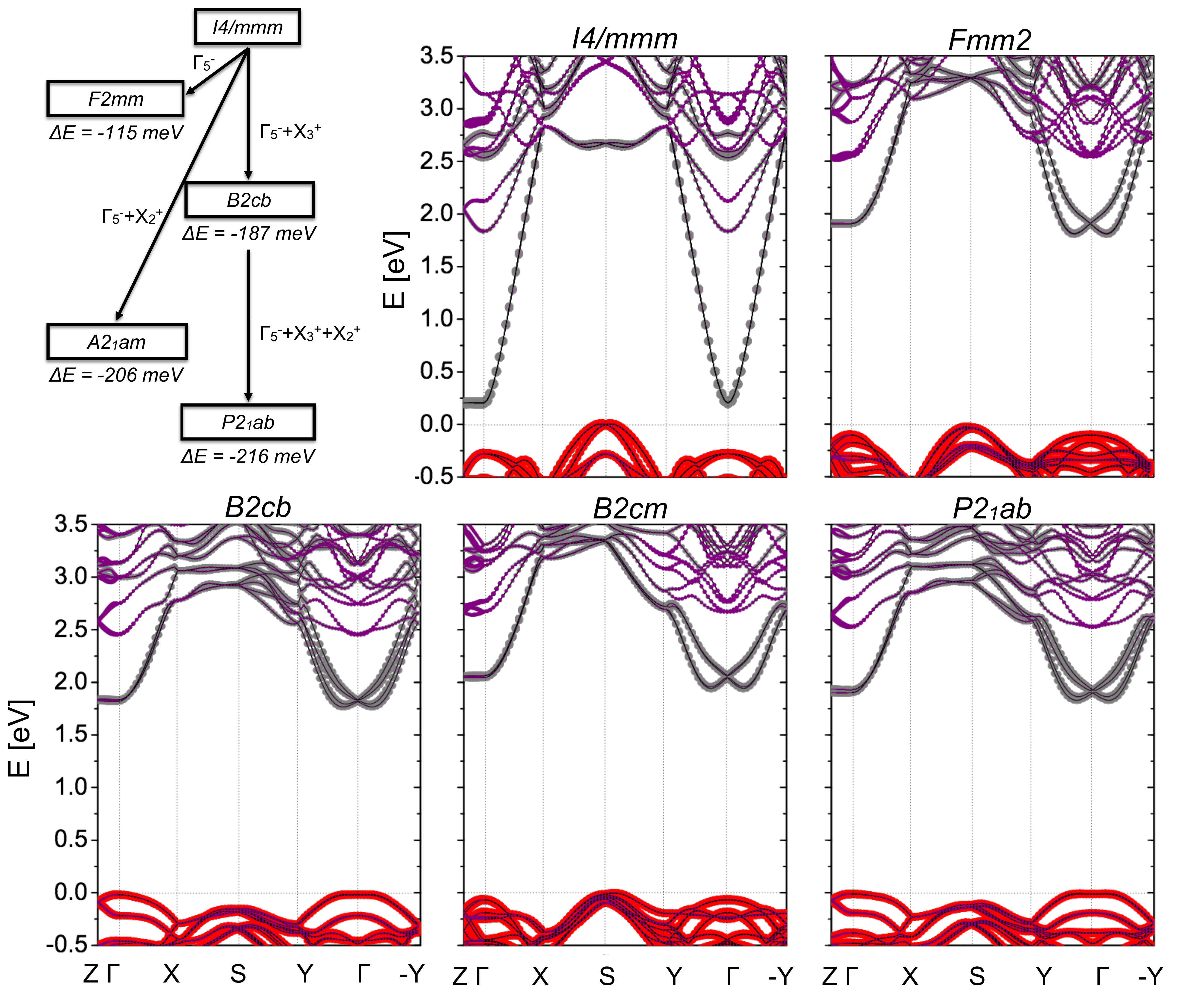}
 \caption{(Color online) Atomically-projected electronic bands dispersions for the reported phases of Bi$_2$WO$_6$ as well as its diagram of the expected phase transitions ($\Delta$ E is the energy difference with respect tio the paraelectric $I4/mmm$). Color code: Grey for W, Purple for Bi and Red for O.}
 \label{S3}
\end{figure}

The atomically projected bands for the different phases of Bi$_2$WO$_6$ are presented in FIG. S\ref{S3}. As expected from the rather ionic character of the compound, the lowest conduction bands are dominated by the W 5d states while the highest valence states have a dominant O 2p character.  The Bi-6p bands lay at much higher energy. 
 
\subsection{C. Role of individual distortions on the RSS }

In a complementary way to Table I comparing the RSS in the relaxed metastable phases of Bi$_2$WO$_6$, in TABLE S \ref{coupling}, we report the evolution  of the Rashba parameter $\alpha_R$ and theoretical bandgap when condensing individually or together the 3 primary distortions connecting the $I4/mmm$ reference to the $P2_1ab$ ground state. As deduced already from Table I, the RSS is essentially produced by the $\Gamma_5^-$ polar distortion, the $c$-axis rotation $X_2^+$ does not have any effect on it and the $a$-axis tilt  $X_3^+$  tends to reduce it strength. Here the volume is fixed to the paraelectric $I4/mmm$ reference to avoid any strain effect (this justifies why the values in TABLE S \ref{coupling} and in TABLE I are slightly different).

\begin{table*}[!tb]
\caption{evolution  of the Rashba parameter $\alpha_R$ (eV$\cdot$\r{A})  and theoretical bandgap (eV) when condensing individually or together the 3 primary distortions ($\Gamma_5^{-}$,  $X_2^{+}$ and $X_3^{+}$) connecting the $I4/mmm$ reference to the $P2_1ab$ ground state. The cell parameters are fixed to those of the  $I4/mmm$ phase.} 
\centering
\begin{center}
\label{coupling}
\begin{tabular}{lllllllll}
\hline
\hline
Modes  & $\alpha_R$ (eV$\cdot$\r{A})   && E$_g$ (eV) \rule[-1ex]{0pt}{3.5ex}  \\
\hline
1) 0.57 $\Gamma_5^{-}$  & 1.07  && 1.63 \rule[-1ex]{0pt}{3.5ex} \\
2)  0.41 $X_2^{+}$  & 0.00   && 0.35 \rule[-1ex]{0pt}{3.5ex} \\
3) 0.72 $X_3^{+}$ & 0.00   && 0.21 \rule[-1ex]{0pt}{3.5ex} \\
4) 0.72 $X_3^{+}$ $\oplus$ 0.41 $X_2^{+}$  &  0.00 && 0.28 \rule[-1ex]{0pt}{3.5ex} \\
5)  0.57 $\Gamma_5^{-}$ $\oplus$ 0.41 $X_2^{+}$ & 1.08   && 1.64 \rule[-1ex]{0pt}{3.5ex} \\
6) 0.57 $\Gamma_5^{-}$ $\oplus$ 0.72 $X_3^{+}$  & 0.83   && 1.25 \rule[-1ex]{0pt}{3.5ex} \\
7) 0.57 $\Gamma_5^{-}$ $\oplus$ 0.72 $X_3^{+}$ $\oplus$ 0.41 $X_2^{+}$  & 0.90 && 1.20 \rule[-1ex]{0pt}{3.5ex} \\
\hline
\hline
\end{tabular}
\end{center}
\end{table*}

%
%

\begin{center}
\section{IV. Other Aurivillius materials}
\end{center}

\begin{figure}[h!tb]
 \centering
 \includegraphics [width=14cm,keepaspectratio=true]{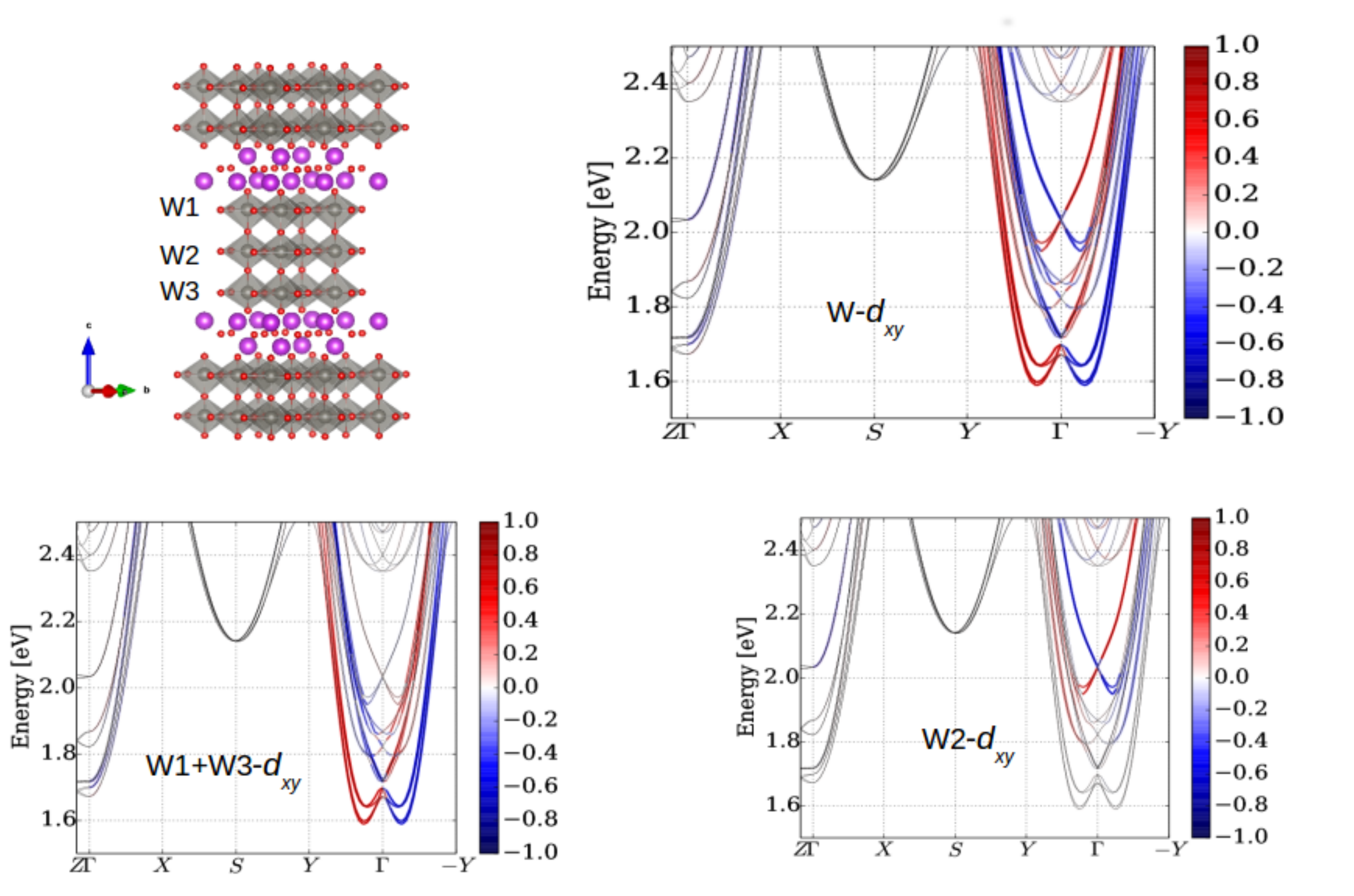}
 \caption{(Color online) Spin-projected band for Bi$_2$W$_3$O$_{12}$.}
 \label{S4}
\end{figure}

\begin{table*}[htb]
\caption{Rashba parameter in various Aurivillius phases and materials.} 
\label{aurivillius}
\begin{center}
\squeezetable
\begin{tabular}{llllclllc}
\hline
\hline
Material  &&&& Phase &&&& $\alpha_R$ (eV$\cdot$\r{A}) \\
\hline
 Bi$_2$WO$_6$ &&&& $P2_1ab$   &&&& 0.88 \\
 Bi$_2$W$_2$O$_9$ &&&& $Cmc2_1$   &&&& 1.58 \\
 Bi$_2$W$_3$O$_{12}$ &&&& $Fmm2$  &&&& 1.61 \\
 SrBi$_2$Ta$_2$O$_9$ &&&& $Cmc2_1$  &&&& 0.30 \\   
 Bi$_4$Ti$_3$O$_{12}$ &&&& $B2cb$  &&&& 0.49 \\ 
\hline
\hline
\end{tabular}
\end{center}
\end{table*}

In order to generalize our findings, we computed the band structure and Rashba spin splitting for Bi$_2$W$_2$O$_9$,  Bi$_2$W$_3$O$_{12}$, SrBi$_2$Ta$_2$O$_9$, and Bi$_4$Ti$_3$O$_{12}$ Aurivillius compounds. The values of $\alpha_R$ are given in TABLE S \ref{aurivillius}. In FIG. S9, we  show also the band structure of Bi$_2$W$_3$O$_{12}$: it is interesting to note that the splitting at the conduction band bottom is due to $d_{x'y'}$ orbitals of W$_1$ and W$_3$, the interface tungsten atoms that are linked to the Bi$_2$O$_2$ layers, while the $d_{x'y'}$ orbital of the central W$_2$ atom is located at higher energy.

\bibliography{library-merge}



\end{document}